\def\lesssim{\mathrel{\hbox{\rlap{\hbox{\lower4pt\hbox{$\sim$}}}\hbox{$<$}}}}

\def\gtrsim{\mathrel{\hbox{\rlap{\hbox{\lower4pt\hbox{$\sim$}}}\hbox{$>$}}}}

\def\teff{$T_{\rm eff}$}

\def\lteff{log $T_{\rm eff}$~}

\def\ll_lsun{log$({L/\rm L_{\odot}})$~}

\def\masa_msun{$M/ \rm M_{\odot}$~}

\def\m_mstar{$M/M_{*}$~}

\documentclass{aa}
\usepackage{graphicx}
        
\begin{document}

\title{The  formation and  evolution  of  hydrogen-deficient  post-AGB 
white dwarfs:  the emerging chemical profile and  the expectations for
the PG1159-DB-DQ evolutionary connection}

\author{L. G. Althaus$^1$\thanks{Member of the Carrera del Investigador
        Cient\'{\i}fico y Tecnol\'ogico  and IALP, CONICET / FCAG-UNLP, 
        Argentina.},
        A. M. Serenelli$^2$,
        J. A. Panei$^3$\thanks{Fellow of CONICET, Argentina.},
        A.  H.  C\'orsico$^3$\thanks{Member  of   the  Carrera     del 
        Investigador   Cient\'{\i}fico   y   Tecnol\'ogico,   CONICET, 
        Argentina.},
        E. Garc\'{\i}a-Berro$^{1,5}$, \\ and
        C. G. Sc\'occola$^3$\thanks{Fellow of CONICET, Argentina.} }
\offprints{L. G. Althaus}

\institute{
$^1$ Departament de F\'\i sica Aplicada, Universitat Polit\`ecnica 
de Catalunya, Av. del Canal Ol\'\i mpic, s/n, 08860, Castelldefels, 
Barcelona, Spain\\
$^2$ Institute for Advanced Study, School of Natural Sciences, Einstein
Drive, Princeton, NJ, 08540, USA\\
$^3$   Facultad   de   Ciencias  Astron\'omicas   y   Geof\'{\i}sicas,
Universidad Nacional de La Plata,  Paseo del Bosque S/N, (B1900FWA) La
Plata, Argentina.\\
$^4$ Instituto  de Astrof\'{\i}sica La Plata, IALP, CONICET-UNLP\\
$^5$  Institut d'Estudis  Espacials  de Catalunya,  Ed. Nexus,  c/Gran
Capit\`a 2, 08034, Barcelona, Spain.\\
\email{althaus,    panei,    acorsico,    cscoccola@fcaglp.unlp.edu.ar, 
       aldos@ias.edu, garcia@fa.upc.es} }

\date{Received; accepted}

\abstract{This   paper is  designed  to  explore  the  formation  and 
evolution of  hydrogen-deficient post-AGB white dwarfs.   To this end,
we compute  the complete evolution  of an initially $2.7  \, M_{\sun}$
star from the zero-age main sequence through the thermally pulsing and
mass-loss phases  to the white  dwarf stage.  Particular  attention is
given to the chemical abundance changes during the whole evolution.  A
time-dependent  scheme  for the  simultaneous  treatment of  abundance
changes  caused  by nuclear  reactions,  diffusive overshooting,  salt
fingers and convection is considered. We employed the double-diffusive
mixing-length  theory  of   convection  for  fluids  with  composition
gradients  (Grossman  \&  Taam  1996).   The study  can  therefore  be
considered  as a  test of  its performance  in low-mass  stars.  Also,
time-dependent  element diffusion  for multicomponent  gases  is taken
into  account  during the  white  dwarf  evolution.  The  evolutionary
stages corresponding  to the  last helium thermal  pulse on  the early
white-dwarf cooling  branch and  the following born-again  episode are
carefully  explored. Relevant aspects  for PG1159  stars and  DB white
dwarf evolution  are studied  in the frame  of these  new evolutionary
models  that  take  into  account  the  history  of  the  white  dwarf
progenitor.  The scope  of the calculations is extended  to the domain
of the  helium-rich, carbon-contaminated DQ white dwarfs  with the aim
of   exploring  the  plausibility   of  the   evolutionary  connection
PG1159-DB-DQ.  In this regard, the implications for the double-layered
chemical  structure  in pulsating  DB  white  dwarfs is  investigated.
Another  aspect of  the  investigation concerned  the consequences  of
mass-loss  episodes   during  the   PG1159  stage  for   the  chemical
stratification of the outer layer of DB and DQ white dwarfs.
 }

\authorrunning{Althaus et al.}

\titlerunning{The  formation   and   evolution  of  hydrogen-deficient 
post-AGB white dwarfs.}

\maketitle


\section{Introduction}

White-dwarf stars  with helium-rich atmospheres,  commonly referred to
as  DB white  dwarfs, comprise  about 20\%  of the  total  white dwarf
population.  Most of these stars  are widely believed to be the result
of a born-again episode, which  is considered to be the most promising
scenario    to   explain    the   existence    of   hydrogen-deficient
post-asymptotic-giant-branch (post-AGB)  stars --- see,  for instance,
Fujimoto  (1977), Sch\"onberner  (1979) and  Iben et  al.   (1983) for
earlier  references. Other possible  channels that  could lead  to the
formation of DB  white dwarfs would involve the  class of helium-rich,
supergiant R~Coronae Borealis (RCrB) stars, probably linked to the hot
hydrogen-deficient O(He)  stars, and the  hydrogen-poor, carbon-normal
star  evolution  directly from  the  extended  horizontal branch  (AGB
manqu\'e  stars  such  as  SdB  stars)  into  the  white  dwarf  state
(Sch\"onberner 1996; Werner 2001).  Within the born-again scenario, 
a very late helium-shell flash is experienced by a white-dwarf remnant 
during its 
early cooling phase after hydrogen  burning has almost ceased.  At the
beginning  of  this  thermal  pulse,  most of  the  residual  hydrogen
envelope  is   engulfed  by  the  helium-flash   convection  zone  and
completely burnt.  The  star is then forced to  evolve rapidly back to
the AGB  and finally into  the central star  of a planetary  nebula at
high effective  temperatures (\teff) but now  as a hydrogen-deficient,
quiescent helium-burning object. Such  objects are expected to exhibit
surface layers  that are substantially  enriched with the  products of
helium burning, particularly carbon.

Important observed examples of these hydrogen-defi-cient post-AGB stars
are the very hot PG1159 and their probable progenitors, the Wolf-Rayet
type  central   stars  of  planetary  nebulae   having  spectral  type
[WC]\footnote{Not  to  be  confused  with massive  Wolf-Rayet  stars.}
(Koesterke  \& Hamann  1997;  Dreizler \&  Heber  1998; Werner  2001).
Indeed,  spectroscopic  analyses  have  revealed that  most  of  these
post-AGB  stars are characterized  by hydrogen-deficient  and helium-,
carbon-  and  oxygen-rich  surface  abundances.   In  particular,  the
appreciable abundance of oxygen in  the atmospheres of these stars has
been successfully explained  by Herwig et al.  (1999)  on the basis of
evolutionary calculations of the born-again scenario that incorporate
convective overshoot.

Strong  observational  evidence suggests  that  PG1159  stars are  the
direct predecessors of the majority  of helium-rich DO stars which are
the  hot and  immediate progenitors  of DB  white dwarfs  (Dreizler \&
Werner  1996; Dreizler \&  Heber 1998).  Theoretical evidence  for the
existence of an evolutionary link between PG1159 stars and most of the
DO white  dwarfs has been presented  by Unglaub \& Bues  (2000) on the
basis of diffusion  calculations with mass loss for  hot white dwarfs.
In   addition,   evolutionary   calculations   taking   into   account
time-dependent element diffusion (Dehner  \& Kawaler 1995; Gautschy \&
Althaus 2002) have  shown that, as a result  of gravitational settling
of carbon and oxygen,  the PG1159-like initial chemical stratification
of  a pre-white dwarf  evolves into  a superficially  helium dominated
double-layered chemical structure when the domain of the pulsating DBs
is  reached. In fact,  two different  chemical transition  zones would
characterize the  envelope of the PG1159 descendants:  a still uniform
intershell  region which  is rich  in helium,  carbon and  oxygen; the
relics of the short-lived mixing episode that occurred during the last
helium  thermal  pulse, and  an  overlying  pure  helium mantle  which
thickens  as  cooling  proceeds.    These  works  clearly  foster  the
plausibility of an evolutionary  connection between most of the PG1159
and DB stars.

The  shape of  the outer  layer chemical  profile is  a matter  of the
utmost importance  as far  as the pulsational  properties of  DB white
dwarfs  is concerned.  In  fact, the  presence of  a diffusion-induced
double-layered  chemical  structure  has  been shown  by  Fontaine  \&
Brassard  (2002)  to  have  strong implications  for  the  theoretical
pulsational  spectrum of  these  stars.  According  to these  authors,
asteroseismological inference  about core  composition of DBs  and the
$^{12}$C($\alpha,\gamma)^{16}$O reaction  rate based on single-layered
DB  models (with  a pure  helium envelope  atop a  carbon-oxygen core)
should  be   taken  with   a  pinch  of   salt.   More   recently,  DB
asteroseismological  fittings  incorporating  both the  double-layered
envelope feature  expected from time-dependent  diffusion calculations
and adjustable carbon-oxygen cores  have been presented by Metcalfe et
al. (2003)  for a  wide range of  helium contents and  stellar masses.
Despite  their  models  yielding  significantly  better  fits  to  the
observations, the  derived stellar  parameters for some  fittings lead
them   to  conclude   that  double-layered   models   with  adjustable
carbon-oxygen  cores may not  be entirely  appropriate to  explain the
observations.

The  possibility that  PG1159 stars  could eventually  evolve  into DB
white  dwarfs  that  are  characterized  by envelopes  with  a  single
composition transition  zone has recently been explored  by Althaus \&
C\'orsico (2004).   Indeed, on the basis  of evolutionary calculations
that   incorporate  time-dependent   element  diffusion,   Althaus  \&
C\'orsico (2004) have shown that if the helium content in PG1159 stars
is smaller than $\approx 10^{-3} M_*$, the double-layered structure is
expected to become single-layered  by the time evolution has proceeded
to the  domain of the variable  DBs.  Although the  small quoted value
for  the helium content  is difficult  to reconcile  with evolutionary
calculations for  the formation of  hydrogen-deficient post-AGB, which
predict the total helium mass left in the star to be about $M_{\rm He}
\approx  10^{-2} M_*$ (Herwig  et al.  1999), the  existence of PG1159 
stars with low helium content cannot be discarded.  In fact, mass-loss
rates ranging from  $10^{-7}$ to $10^{-8}\, M_{\sun}$/yr  are observed
in many  luminous PG1159 stars.   In addition, tentative  evidence for
the  persistence of  mass-loss  rates within  the  range $10^{-7}$  to
$10^{-10} \, M_{\sun}$/yr down to  the domain of hot helium-rich white
dwarfs  has  been presented  (Werner  2001).   The  existence of  such
mass-loss rates would  imply that most of the  helium-rich envelope of
DB progenitors could be substantially reduced during the time interval
mass loss would be operative.
It  is worth  mentioning that  the existence  of PG1159  stars  with a
helium content as  low as $10^{-3} \, M_{\sun}$  has been suggested by
asteroseismology in at least one of these stars with a stellar mass of
$0.6  \,  M_{\sun}$  (Kawaler  \&  Bradley 1994),  thus  implying  the
occurrence  of modest  mass-loss during  the evolution  to  the PG1159
phase.

In this work we explore some  relevant aspects for the evolution of DB
white  dwarfs  on the  basis  of  new  evolutionary calculations  that
account  for   a  complete   and  self-consistent  treatment   of  the
evolutionary  stages   prior  to   the  white  dwarf   formation.   We
concentrate on  DB white dwarfs  resulting from a  born-again episode.
Specifically, we follow the evolution of an initially $2.7\, M_{\sun}$
star from the zero-age main sequence through the thermally pulsing and
mass-loss  phases  on  the  AGB   to  the  white  dwarf  regime.   The
evolutionary stages  corresponding to  the almost complete  burning of
protons following  the occurrence of  the very late thermal  pulse and
the ensuing  born-again episode are carefully  explored.  Attention is
paid to  the abundance  changes along the  whole evolution,  which are
described  by means of  a time-dependent  scheme for  the simultaneous
treatment of nuclear evolution and mixing processes due to convection,
salt fingers and diffusive  overshoot.  We emphasize in particular the
role  of time-dependent  element diffusion  in the  chemical abundance
distribution  in  the  white-dwarf  regime. We  also  investigate  the
influence of  mass-loss episodes during  the PG1159 and DO  phases for
the chemical  stratification of pulsating DBs. Finally,  we extend the
scope  of   our  calculations  to   the  domain  of   the  helium-rich
carbon-contaminated DQ  white dwarfs, the  supposed cooler descendants
of DBs. The plausibility  of the  evolutionary  connection PG1159-DB-DQ
(Fontaine \& Brassard  2002) is assessed in  the framework of our
new evolutionary models.

As far as we  are aware, this is the first time  that the evolution of
hydrogen-deficient white dwarfs is performed consistently on the basis
of a  complete and detailed  treatment of the physical  processes that
lead to the formation of such  stars. At this point it is important to
note that we think that a re-examination of the pulsational properties
of variable DB  white dwarfs deserves to be performed  in the frame of
the new  evolutionary models presented  in this work.   However, being
this issue important it is also  true that this would carry us too far
afield.   The paper is  organized as  follows.  The  following section
contains  details   on  the  main  physical  inputs   to  the  models,
particularly  regarding  the   treatment  of  the  chemical  abundance
changes.  In Sect.   3 we present the evolutionary  results. There, we
elaborate  on  the main  aspects  of  the  pre-white dwarf  evolution,
particularly during  the born-again  phase and the  attendant chemical
changes.   In   the  same  section,  we  also   describe  the  results
corresponding to the  PG1159 and white-dwarf regimes.  In  Sect. 4, we
discuss the implications of our  results for the white dwarf evolution
as well as the role played by mass-loss episodes.  Finally, Sect. 5 is
devoted to discuss some concluding remarks.


\section{Input physics and evolutionary sequences}

The calculations presented in this  work have been done using a Henyey
type stellar  evolution code.  In  particular, we employed  the LPCODE
evolutionary  code  that  is  specifically  designed  to  compute  the
formation and  evolution of white  dwarf stars, following a  star from
the  main sequence  through the  thermal pulses  and  post-AGB phases.
Except for minor modifications, the code is essentially that described
at length  in Althaus  et al.  (2003)  and references  therein. LPCODE
uses  OPAL  radiative  opacities  (including carbon-  and  oxygen-rich
compositions) for  different metallicities (Iglesias  \& Rogers 1996),
complemented, at  low temperatures, with the  molecular opacities from
Alexander \&  Ferguson (1994).  Opacities  for different metallicities
are  required  in particular  during  the  white  dwarf regime,  where
metallicity gradients induced by gravitational settling develop in the
envelope of  such stars.  High-density conductive  opacities are those
of  Itoh  et al.   (1994)  and  the  references cited  there,  whereas
neutrino  emission  rates are  those  of  Itoh  (1997) and  references
therein.  The  equation of state  for the low-density  regime includes
partial  ionization for  hydrogen and  helium  compositions, radiation
pressure  and  ionic  contributions.   For  the  high-density  regime,
partially  degenerate  electrons  and  Coulomb interactions  are  also
considered.  During the born-again  episode, the remnant star develops
surface layers  rich in helium, carbon  and oxygen.  In  that case, we
employ an ideal equation of state that includes partial ionization for
any mixture  of the  three chemical species.   Finally, for  the white
dwarf  regime, we  considered an  updated version  of the  equation of
state of Magni \& Mazzitelli (1979).

The  nuclear network employed  in LPCODE  accounts explicitly  for the
following 16 elements: $^{1}$H, $^{2}$H, $^{3}$He, $^{4}$He, $^{7}$Li,
$^{7}$Be, $^{12}$C, $^{13}$C,  $^{14}$N, $^{15}$N, $^{16}$O, $^{17}$O,
$^{18}$O, $^{19}$F, $^{20}$Ne and $^{22}$Ne.  In addition, we consider
34   thermonuclear   reaction   rates   to   describe   the   hydrogen
(proton-proton chain and CNO  bi-cycle) and helium burning, and carbon
ignition. Specifically, for hydrogen burning we consider: \\

\noindent $p + p \rightarrow \ ^{2}$H + $e^+ + \nu$ 

\noindent $p + p  + e^- \rightarrow \ ^{2}$H + $\nu$

\noindent $^{2}$H$\ +\ p  \rightarrow \ ^{3}$He + $\gamma$ 

\noindent $^{3}$He$\ +\ ^{3}$He$  \rightarrow \ \alpha$ + $2p$ 

\noindent $^{3}$He$\ +\ \alpha  \rightarrow \ ^{7}$Be + $\gamma$ 

\noindent $^{3}$He$\ +\ p \rightarrow \ ^{4}$He + $\gamma$  

\noindent $^{7}$Be$\ +\ e^-  \rightarrow \ ^{7}$Li + $\nu$ 

\noindent $^{7}$Li$\ +\ p  \rightarrow \ 2\alpha$ 

\noindent $^{7}$Be$\ +\ p  \rightarrow \ 2\alpha$ 

\noindent $^{12}$C$\ +\ p  \rightarrow \ ^{13}$C + $ e^+ + \nu$ 

\noindent $^{13}$C$\ +\ p  \rightarrow \ ^{14}$N + $\gamma$ 

\noindent $^{14}$N$\ +\ p  \rightarrow \ ^{15}$N + $ e^+ + \nu$

\noindent $^{15}$N$\ +\ p  \rightarrow \ ^{12}$C + $\alpha$ 

\noindent $^{15}$N$\ +\ p  \rightarrow \ ^{16}$O + $\gamma$ 

\noindent $^{16}$O$\ +\ p  \rightarrow \ ^{17}$O + $ e^+ + \nu$ 

\noindent $^{17}$O$\ +\ p  \rightarrow \ ^{18}$O + $ e^+ + \nu$ 

\noindent $^{17}$O$\ +\ p  \rightarrow \ ^{14}$N + $\alpha$ 

\noindent $^{18}$O$\ +\ p  \rightarrow \ ^{15}$N + $\alpha$ 

\noindent $^{18}$O$\ +\ p  \rightarrow \ ^{19}$F + $\gamma$ 

\noindent $^{19}$F$\ +\ p  \rightarrow \ ^{16}$O + $\alpha$ 

\noindent $^{19}$F$\ +\ p  \rightarrow \ ^{20}$Ne + $\gamma$ \\

\noindent For helium burning, the reaction rates taken into account are: \\

\noindent $ 3\alpha  \rightarrow \ ^{12}$C + $\gamma$ 

\noindent $^{12}$C$\ +\ \alpha  \rightarrow \ ^{16}$O + $\gamma$ 

\noindent $^{13}$C$\ +\ \alpha  \rightarrow \ ^{16}$O + $n$ 

\noindent $^{14}$N$\ +\ \alpha  \rightarrow \ ^{18}$O + $ e^+ + \nu$ 

\noindent $^{15}$N$\ +\ \alpha  \rightarrow \ ^{19}$F + $\gamma$   

\noindent $^{16}$O$\ +\ \alpha  \rightarrow \ ^{20}$Ne + $\gamma$ 

\noindent $^{17}$O$\ +\ \alpha  \rightarrow \ ^{20}$Ne + $n$

\noindent $^{18}$O$\ +\ \alpha  \rightarrow \ ^{22}$Ne + $\gamma$ 
               
\noindent $^{20}$Ne$\ +\ \alpha  \rightarrow \ ^{24}$Mg + $\gamma$ 

\noindent $^{22}$Ne$\ +\ \alpha  \rightarrow \ ^{25}$Mg + $n$ 

\noindent $^{22}$Ne$\ +\ \alpha  \rightarrow \ ^{26}$Mg + $\gamma$ \\

\noindent This set of nuclear reactions allows us to follow in detail the main
nucleosynthesis  occurring  during   the  thermally  pulsing  AGB  and
born-again phases. Finally, the reactions for carbon burning are: \\

\noindent $^{12}$C$\ +\ ^{12}$C$  \rightarrow \ ^{20}$Ne + $\alpha$ 

\noindent $^{12}$C$\ +\ ^{12}$C$  \rightarrow \ ^{24}$Mg + $\gamma$ \\

Nuclear  reaction rates  are  taken from  Caughlan  \& Fowler  (1988),
except  for the  reactions  $^{15}$N$(p, \gamma)^{16}$O,  $^{15}$N$(p,
\alpha)^{12}$C,      $^{18}$O$(p,     \alpha)^{15}$N,     $^{18}$O$(p,
\gamma)^{19}$F,                        $^{12}$C$(\alpha,\gamma)^{16}$O,
$^{16}$O$(\alpha,\gamma)^{20}$Ne,           $^{13}$C$(\alpha,n)^{16}$O,
$^{18}$O$(\alpha,\gamma)^{22}$Ne,   $^{22}$Ne$(\alpha,n)^{25}$Mg   and
$^{22}$Ne$(\alpha,\gamma)^{26}$Mg,  which  are  taken from  Angulo  et
al.  (1999).   In particular,  it  is  important  to realize  for  the
purposes of our work that the $^{12}$C$(\alpha,\gamma)^{16}$O reaction
rate given by Angulo et al. (1999)  is about twice as large as that of
Caughlan \& Fowler (1988).

An  important aspect  of the  present study  is the  treatment  of the
abundance  changes throughout all  the different  evolutionary phases.
In particular,  during some short-lived  phases of the  evolution, for
instance during the born-again  episode and also the thermally pulsing
AGB phase, the nuclear time-scale of some reactions becomes comparable
to the convective mixing time-scale. In this case the instantaneous 
mixing
approximation  turns out  to be  completely inadequate  for addressing
chemical  mixing  in  convective  regions.  A  more  physically  sound
chemical  evolution  scheme  than  instantaneous mixing  is  therefore
required.  In  this work, we  have considered a  time-dependent scheme
for the  simultaneous treatment of chemical changes  caused by nuclear
burning  and mixing  processes.  Specifically,  abundance  changes are
described by the set of equations

\begin{equation} \label{ec1}
\left( \frac{d \vec{Y}}{dt} \right) = 
\left( \frac{\partial \vec{Y}}{\partial t} \right)_{\rm nuc} +
\frac{\partial}{\partial M_r} \left[ (4\pi r^2 \rho)^2 D 
\frac{\partial \vec{Y}}{\partial M_r}\right], 
\end{equation} 

\noindent with  $\vec{Y}$  being  the  vector  containing  the  number 
fraction  of all  considered  elements.  Details  about the  numerical
procedure  for  solving this  equation  can  be  found in  Althaus  et
al.  (2003).   Here,  mixing  due  to  convection,  salt  fingers  and
overshoot  is treated  as  a  diffusion process  (second  term of  Eq.
\ref{ec1}).  The  efficiency of  convective and salt-finger  mixing is
described  by   appropriate  diffusion  coefficients   $D$  which  are
specified by our treatment of convection. In particular, we considered
the extended mixing-length theory  (MLT) of convection for fluids with
composition  gradients developed  by Grossman  et al.   (1993)  in its
local  approximation  as given  by  Grossman  \&  Taam (1996).   These
authors  have   developed  the  non-linear   MLT  of  double-diffusive
convection\footnote{The  effects  of   both  thermal  and  composition
gradients  determine the  stability of  the fluid.}   that  applies in
convective,  semiconvective and  salt-finger instability  regimes.  In
this work,  the mixing length free  parameter $\alpha$ is  taken to be
1.5.  To our knowledge, our  work constitutes the first application of
the extended  MLT to full evolutionary calculation  of low-mass stars.
The first term  of Eq. (\ref{ec1}) gives the  abundance changes due to
thermonuclear  reactions.   This  term  is  linearized  following  the
implicit  scheme  developed  by  Arnett  \&  Truran  (1969).   In  our
treatment of abundance changes,  nuclear evolution is fully coupled to
the current composition change due to the various mixing processes ---
see  Althaus  et  al.   (2003)  for additional  details.   In  LPCODE,
abundance changes  are performed for  the 16 elements  described above
after  the   convergence  of  each  stellar  model   (and  not  during
iterations). 

Mixing  episodes  beyond  what   is  predicted  by  the  Schwarzschild
criterion for convective stability  strongly affect the inner chemical
profile  of white dwarf  progenitors.  The  occurrence of  such mixing
episodes,  particularly core  overshooting  and/or semiconvection,  is
suggested  by both theoretical  and observational  evidence. Recently,
Straniero et  al.  (2003)  have presented an  assessment of  the inner
chemical  abundances in  a $3\,  M_{\sun}$ model  star  resulting from
different  mixing processes occurring  during the  late stage  of core
helium burning phase.  In  particular, they conclude that models which
incorporate semiconvection or  a moderate overshoot applied
to  core and  convective  shells,  predict a  sharp  variation of  the
chemical composition  in the carbon-oxygen  core.  On the  other hand,
the presence of overshooting  below the convective envelope during the
thermal pulses has been shown by  Herwig et al.  (1997) to yield third
dredge-up and  carbon-rich AGB stars  for relatively low  initial mass
progenitors --- see  also Ventura et al. (1999)  and Mazzitelli et al.
(1999).  In  addition, overshooting below  the helium-flash convection
zone during the  thermally pulsing AGB phase gives  rise to intershell
abundances   in    agreement   with   abundance    determinations   in
hydrogen-deficient post-AGB  remnants such as PG1159  stars (Herwig et
al.  1999;  Herwig 2000).   In view of  these considerations,  we have
allowed for some mechanical overshoot  in our work.  In particular, we
have included time-dependent  overshoot mixing during all evolutionary
stages.  Our  scheme for  the changes in  the abundances allows  for a
self-consistent treatment of diffusive overshooting in the presence of
nuclear burning.  We  have considered exponentially decaying diffusive
overshooting  above and  below {\sl  any} formally  convective region,
including  the  convective  core  (main sequence  and  central  helium
burning phases), the external  convective envelope and the short-lived
helium-flash convection zone which  develops during the thermal pulses
and  born-again  episode.   In  particular,  the  expression  for  the
diffusion coefficient in overshoot regions is $D_{\rm OV}= D_{\rm C}\
\exp(-2z/H_{\rm v})$ where $D_{\rm C}$ is the diffusion coefficient at 
the edge of  the convection zone, $z$ is the  radial distance from the
boundary of the  convection zone, $H_{\rm v}= f  H_{\rm P}$, where the
free parameter $f$ is a measure of the extent of the overshoot region,
and  $H_{\rm  P}$ is  the  pressure  scale  height at  the  convective
boundary.  In  this study we  have adopted $f= 0.015$,  which accounts
for  the  observed  width  of  the main  sequence  and  abundances  in
hydrogen-deficient, post-AGB objects (Herwig et al. 1997, 1999; Herwig
2000; Mazzitelli  et al.   1999).  We  would like to  mention here that
the  breathing pulse  instability occurring  towards the  end  of core
helium burning has been suppressed --- see Straniero et al. (2003) for
a recent discussion of this point.

Finally, the  evolution of the chemical  abundance distribution caused
by diffusion  processes during  the white dwarf  regime has  also been
taken into account in this work.  Our time-dependent element diffusion
treatment, based on the formulation for multicomponent gases presented
by  Burgers  (1969), considers  gravitational  settling, chemical  and
thermal  diffusion  but  not  radiative  levitation,  which  has  been
neglected,  for  the  nuclear  species  $^{1}$H,  $^{3}$He,  $^{4}$He,
$^{12}$C,  $^{13}$C,  $^{14}$N,  $^{16}$O  and $^{22}$Ne  (Althaus  et
al. 2003).   Diffusion velocities  are evaluated at  each evolutionary
step.  We  stress that during the  white dwarf regime,  the metal mass
fraction $Z$ in the envelope is not assumed to be fixed, instead it is
specified  consistently   according  to  the   prediction  of  element
diffusion.

Given  the  considerable  load  of  computing  time  demanded  by  our
self-consistent  solution  of  nuclear  evolution  and  time-dependent
mixing, we have  limited ourselves to examining only  one case for the
evolution  for  the progenitor  star.   Specifically,  we compute  the
evolution  of an  initially $2.7\,  M_{\sun}$ stellar  model  from the
zero-age main  sequence all  the way from  the stages of  hydrogen and
helium  burning in the  core up  to the  tip of  the AGB  where helium
thermal  pulses  occur.   A  solar-like  initial  composition  $(Y,Z)=
(0.275,0.02)$  has been adopted  (Anders \&  Grevesse 1989).   We have
considered mass-loss  episodes taking place during the  stages of core
helium  burning and  red  giant branch  following  the usual  Reimers
formulation with $\eta_{\rm R}$=1.  During the thermally pulsing phase
we adopt the mass-loss rates from Bl\"ocker (1995). After experiencing
10 thermal  pulses, the  progenitor departs from  the AGB  and evolves
towards high  effective temperatures. Departure from the  AGB has been
forced to  occur at such an  advanced phase of the  helium shell flash
cycle that the post-AGB remnant undergoes a final thermal pulse during
the early white dwarf cooling phase --- a very late thermal pulse, see
Bl\"ocker (2001) for a review  --- where most of the residual hydrogen
envelope  is burnt.  As  mentioned in  the introduction,  there exists
observational  evidence  that  suggests that  some  hydrogen-deficient
post-AGB  stars could  experience  strong stellar  winds, which  could
reduce the  helium content  in the star  considerably.  To  assess the
influence of such mass-loss episodes for the further evolution we have
considered extreme mass-loss rates of  $10^{-7}$ and  $10^{-8}\, M_{\sun}$/yr
during the PG1159 stage.

The evolutionary calculations have been followed down to the domain of
the helium-rich  carbon-contaminated DQ white  dwarfs with the  aim of
exploring the evolutionary connection  between DQ white dwarfs and the
PG1159 stars.   The evolutionary sequence  from the ZAMS to  the white
dwarf stage  comprises $\approx$ 70000 stellar  models. Stellar models
are divided into about 1500 mesh  points.  The final mass of the white
dwarf  remnant is  $0.5885  \,  M_{\sun}$. Our  models  would also  be
particularly  appropriate  for   pulsational  studies  of  PG1159  and
variable DB  white dwarfs.   We report below  the main results  of our
calculations.


\section{Evolutionary results}

\subsection{White dwarf  progenitor:  evolution  from the ZAMS  to  the
thermally pulsing phase}

\begin{figure*}
\centering
\includegraphics[clip,width=500pt]{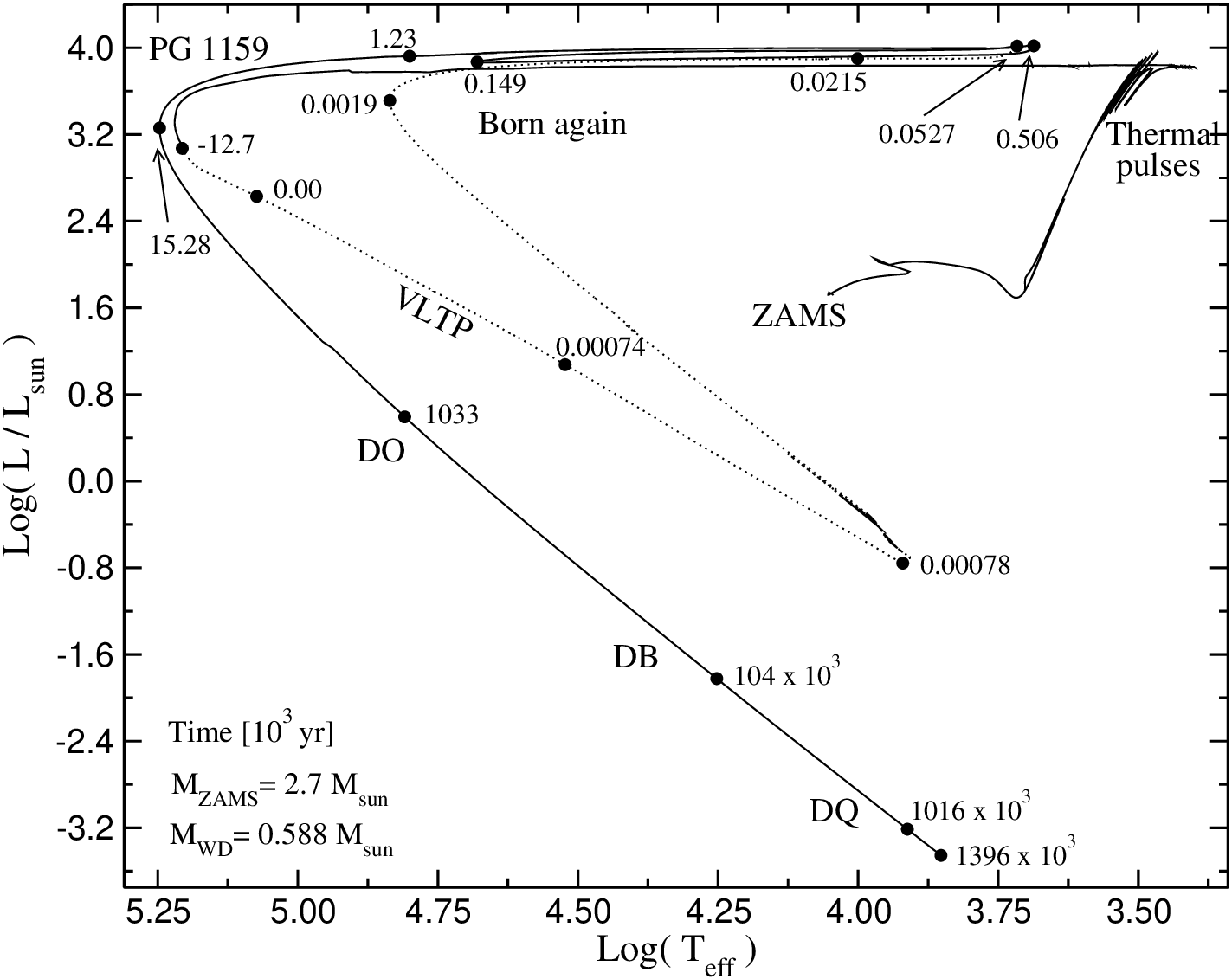}
\caption{Hertzsprung-Russell diagram for the complete evolution of our
$2.7  \, M_{\sun}$  stellar model  from the  ZAMS to  the  white dwarf
domain.  The star experiences a  very last thermal pulse (VLTP) on its
early  cooling phase after  hydrogen burning  has almost  ceased.  The
following evolutionary  stages correspond to the  born-again phase and
are depicted, together with the VLTP stage with  dotted lines. 
The numbers close  to the filled dots
along the  evolutionary track correspond  to the age (in  $10^{3}$ yr)
counted from the occurrence of the peak of the last thermal pulse. The
domain of  the PG1159, DO, DB and  DQ stars are indicated,  as well as
the thermal  pulse and  born-again stages.  As  a result of  mass loss
episodes,  the   stellar  mass  decreases  from  2.7   to  $0.5885  \,
M_{\sun}$.  After  the   born-again  episode,  the  hydrogen-deficient
post-AGB  remnant   experiences  a  second   excursion  towards  lower
temperatures  before  reaching  eventually  its terminal  white  dwarf
track.}
\label{m2.7-hrp.eps}
\end{figure*}

In  Fig. \ref{m2.7-hrp.eps} we  show the  complete Hertzsprung-Russell
(HR)  diagram. Our  numerical simulation  covers all  the evolutionary
phases of  an initially $2.7  \, M_{\sun}$ star  from the ZAMS  to the
domain of the  DQ white dwarfs, including the  stages corresponding to
the helium thermal pulses on the AGB and the born-again episode (shown
as a  dotted line). The age (in  units of 10$^3$ yr)  counted from the
occurrence  of the last  thermal pulse  peak on  the cooling  track is
shown at selected points along the evolutionary track.  The total time
spent in central hydrogen and helium  burning  amounts to $6.57
\times 10^{8}$~yr.  After helium  is exhausted in the core  and during
the following  $1.16 \times 10^{7}$~yr,  the star evolves  towards the
thermally  pulsing phase  on  the AGB.   There,  helium shell  burning
becomes  unstable and the  star experiences  the well  known recurrent
thermal  instability commonly  referred  to as  helium thermal  pulses
(Schwarzschild  \& H\"arm  1965).  After  10 thermal  pulses and  as a
result  of strong  mass  loss, the  remnant  star leaves  the AGB  and
evolves towards high \teff s.  This takes place when the luminosity of
the  star is supported  mostly by  stationary hydrogen  burning.  Mass
loss decreases the stellar mass from 2.7 to $0.5885 \, M_{\sun}$.  Note
that the post-AGB remnant undergoes  a further (last) thermal pulse on
its early white dwarf cooling track shortly after hydrogen burning has
virtually  ceased (Bl\"ocker 2001).   During this  born-again episode,
most  of  the residual  hydrogen  envelope  is  engulfed by  the  deep
helium-flash  convection  zone  and  is completely  burnt.   Evolution
proceeds through these stages very  fast, since it takes only about 30
yr for the remnant to expand from a white dwarf configuration to giant
dimensions.   Note also  that after  the born-again  episode,  the now
hydrogen-deficient  post-AGB remnant  experiences  a second  excursion
towards lower  temperatures (a  double-loop path) before  reaching the
domain of  the PG1159  stars and eventually  its terminal  white dwarf
cooling  track.  It  is worth  noting  that during  these stages,  the
remnant is a quiescent helium-burning object that reaches the point of
maximum \teff\  (the knee in the  HR diagram) in  about 15000~yr after
the occurrence of the last thermal pulse.

\begin{figure}
\centering
\includegraphics[clip,width=250pt]{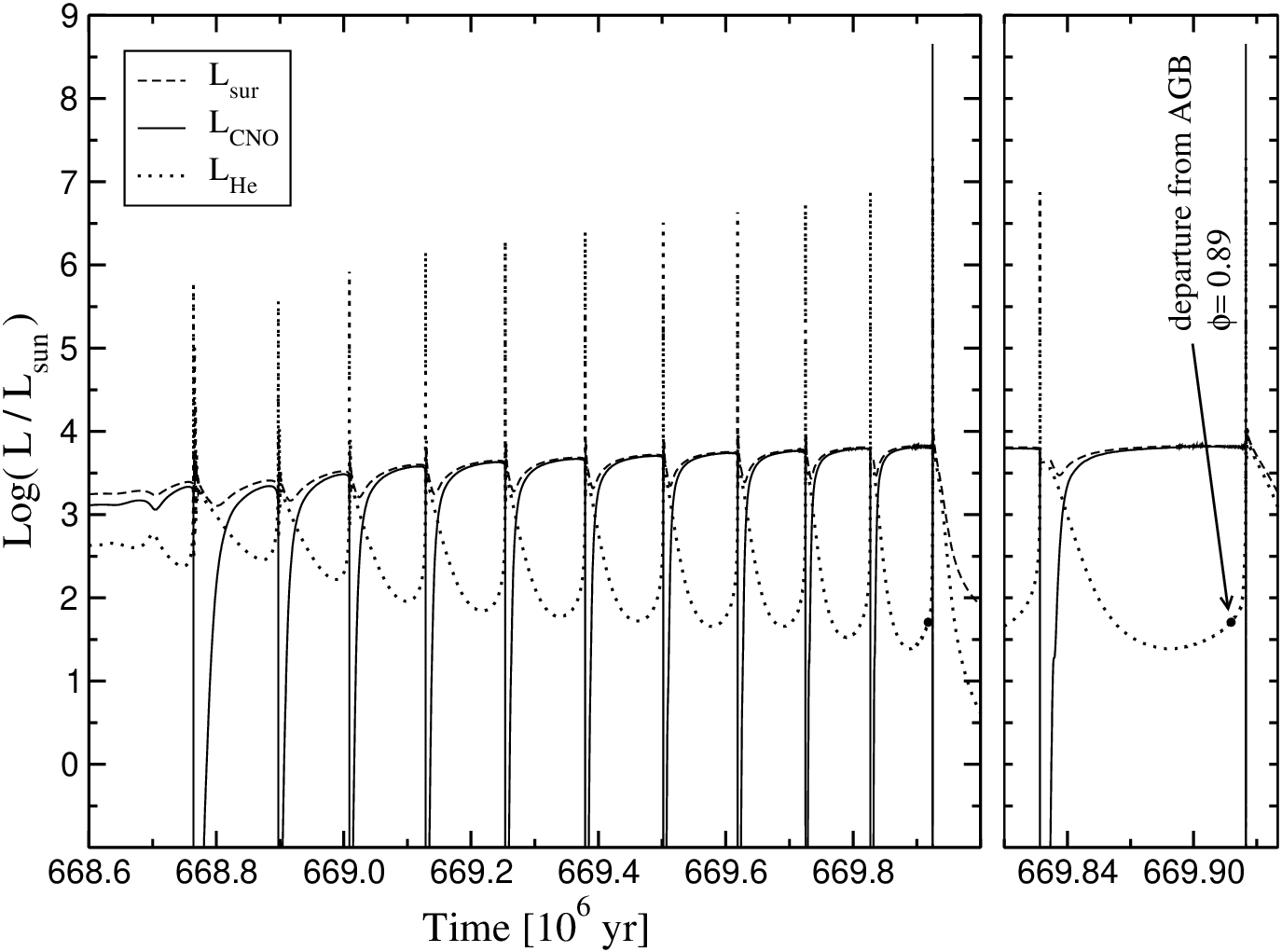}
\caption{The  temporal evolution  of surface luminosity  and hydrogen-  
and helium-burning  luminosities in solar units for  an initially $2.7
\,  M_{\sun}$ star  during the  thermally  pulsing phase  on the  AGB.
After experiencing 10 thermal pulses, the progenitor star departs from
the AGB with  a thermal-pulse cycle phase of 0.89  (denoted by a black
dot).  The  last thermal pulse shown  here takes place  already on the
white  dwarf  cooling  track,  where the  hydrogen-burning  luminosity
exceeds  $10^{8}\, L_{\sun}$.   The right  panel shows  the luminosity
evolution after  the occurrence  of the tenth  pulse. Age is  given in
$10^{6}$~yr counted from the main sequence.}
\label{m2.7pulsos.eps}
\end{figure}

The evolution during the thermally pulsing phase is documented in Fig.
\ref{m2.7pulsos.eps}.  This figure  shows the  time dependence  of the
surface   luminosity   ($L_{\rm   sur}$)   and   the   hydrogen-   and
helium-burning   luminosities  ($L_{\rm   CNO}$   and  $L_{\rm   He}$,
respectively) of our initially $2.7 \, M_{\sun}$ model star, where the
time-scale is given in Myr from  the main sequence (ZAMS).  A total of
10 thermal  pulses with  an interpulse period  of roughly  $1.2 \times
10^{5}$~yr have been  computed before the remnant leaves  the AGB as a
result of the  enhanced mass-loss.  Note that the  helium burning rate
rises very  steeply at the peak  of each pulse, even  during the first
pulses.  In our simulation,
departure from  the AGB takes place  at such an advanced  stage in the
helium shell flash  cycle that the post-AGB remnant  will experience a
last   helium  thermal   pulse   at  high   \teff\   values  ---   see
Fig.  \ref{m2.7-hrp.eps} and  also the  next section.   This situation
corresponds to the last thermal pulse and it is illustrated separately
on  the  right panel  of  Fig. \ref{m2.7pulsos.eps}.   During the  last
pulse, hydrogen-burning luminosity,  due mainly to proton captures
by  $^{12}$C,  reaches about $10^{8.5}\, L_{\sun}$.

\begin{figure}
\centering
\includegraphics[clip,width=250pt]{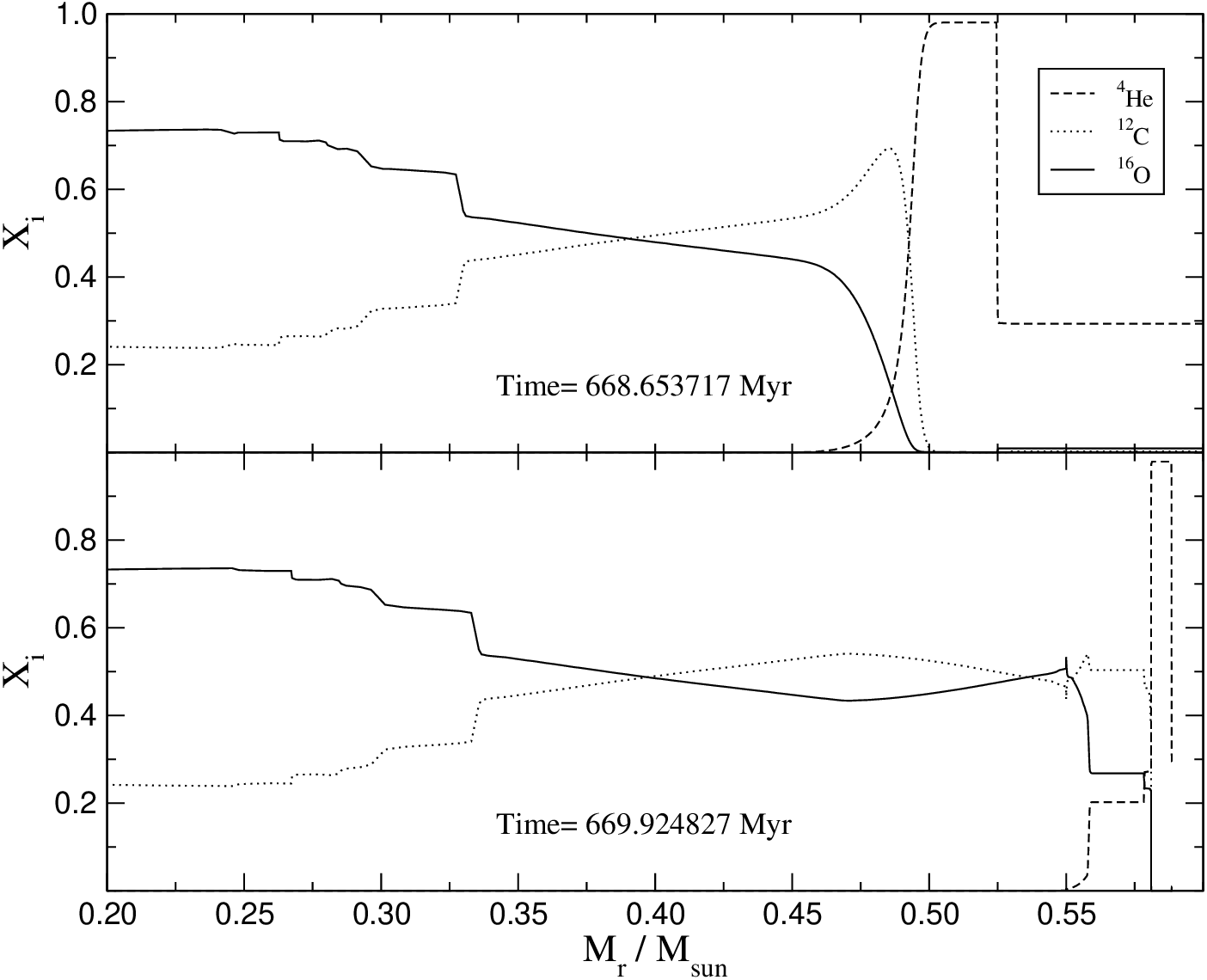}
\caption{Internal $^{4}$He,  $^{12}$C  and $^{16}$O  profiles  for the
$2.7  \,  M_{\sun}$  DB  white  dwarf progenitor  shortly  before  the
occurrence of the first thermal pulse  on the AGB (upper panel) and at
the onset  of the  last thermal pulse  at high \teff\  (bottom panel).
Overshoot  episodes  leave recognizable  features  both  in the  inner
carbon-oxygen profile  and in the intershell  region that are  
particularly rich in oxygen. See text for details.}
\label{Xi_int.eps}
\end{figure}

The  inner carbon, oxygen  and helium  distribution in  the progenitor
star  as  a  function  of   the  mass  coordinate  is  shown  in  Fig.
\ref{Xi_int.eps}. The  upper panel of the figure  depicts the chemical
profile at  the onset  of the thermally  pulsing AGB phase,  while the
bottom panel illustrates  the situation for the beginning  of the last
thermal pulse  at high  \teff.  Some features  of this  figure deserve
comments. In particular, the  carbon and oxygen abundance distribution
within  the  core is  typical  of  situations  in which  extra  mixing
episodes  beyond  the  fully  convective core  during  central  helium
burning are  allowed.  Such  extra mixing episodes,  particularly core
overshooting  and/or semiconvection,  have  a large  influence on  the
carbon and oxygen distribution in  the core of white dwarfs (Straniero
et al.  2003).   In our simulation, the central  oxygen mass abundance
amounts to  0.724 when  the helium convective  core disappears.  It is
worth noting that the shape of the carbon-oxygen profile, particularly
the sharp variation  around $M_r \approx 0.33 \,  M_{\sun}$ induced by
mechanical overshoot, is in  agreement with that reported in Straniero
et  al.   (2003).   The  importance  of such  extra  mixing  episodes,
particularly core  overshoot, for the pulsational properties of massive 
ZZ Ceti stars has recently been  emphasized by Althaus et  al.  (2003)  
and  C\'orsico et  al.   (2004).  Note  that during  the thermally
pulsing  phase,   the  mass   of  the  carbon-oxygen   core  increases
considerably  because the  helium burning  shell moves  outwards.  The
signatures left by overshoot in the chemical profile during this stage
is apparent. In particular,  the overshoot region stretching below the
short-lived  helium-flash convection  zone  at pulse  peak causes  the
intershell  region   below  the  almost  pure  helium   buffer  to  be
substantially  enriched in  oxygen.
Another feature predicted by our calculations is the occurrence of the
third  dredge-up during  the  thermally pulsing  phase.   We will  not
discuss this aspect here --- see Herwig (2000) for details --- suffice
it to say  that as a result of envelope  overshoot, stretching down to
the underlying carbon-rich layers, large amounts of carbon are dredged
up to the surface after each thermal pulse.  Indeed, the carbon-oxygen
ratio  increases  from  $\approx$  0.25 when  entering  the  thermally
pulsing phase to  $\approx$ 0.34 by the time  the remnant departs from
the AGB.  It is  important to point  out at  this point that  we have
checked  the  formation  of  a   carbon  star  in  our  simulation  by
suppressing  mass loss.   We find  in that  case that  a carbon-oxygen
ratio larger than  1 occurs after the occurrence  of the 15$^{\rm th}$
pulse.

\begin{figure*}
\centering
\includegraphics[clip,width=500pt]{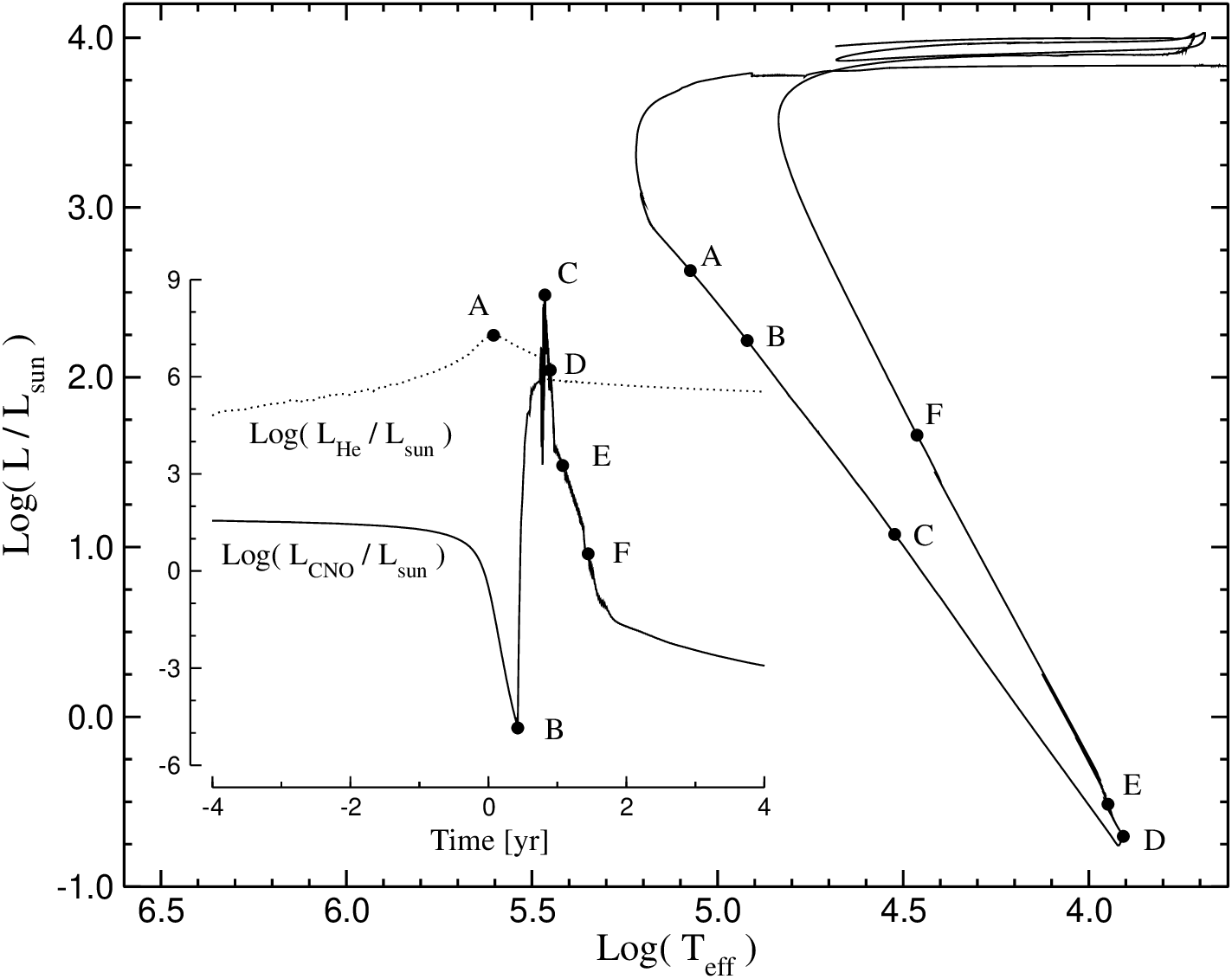}
\caption{Hertzsprung-Russell   diagram  for  the  evolutionary  stages 
following the onset of the  last helium thermal pulse that takes place
at high  effective temperatures (born-again episode)  for our post-AGB
remnant of $0.5885  \, M_{\sun}$. Note that the  remnant experiences a
second excursion (double loop) to the red giant region before evolving
into  its  final  white  dwarf  state. The  left  inset  displays  the
time-dependence  of hydrogen-  and helium-burning  luminosities (solid
and  dotted lines,  respectively) expressed  in solar  units. Selected
stages are labeled by letters  along the curves. In particular, points
A  and C correspond  to the  moment of  maximum helium-  and hydrogen-
burning luminosities,  respectively.  At point B, protons  begin to be
engulfed  and burnt  by the  outward-growing,  helium-flash convection
zone.   Between  points  E  and  F, hydrogen  burning  becomes  almost
extinguished.  The  hydrogen content remaining in the  star amounts to
$1.3 \times 10^{-8}\, M_{\sun}$. Note the extremely short evolutionary
time-scales characterizing this phase.}
\label{m2.7-hr-pfp.eps}
\end{figure*}

Finally, another  prediction of our  calculations is the  formation of
$^{13}$C- and $^{14}$N- pockets at the base of the helium buffer after
the end of  the dredge-up phase.  Indeed, during  the third dredge-up,
diffusive overshoot has  formed a small region in  which hydrogen from
the  envelope and  carbon  (resulting from  helium  burning) from  the
intershell region coexist in  appreciable abundances. When this region
heats up enough, hydrogen reignites  giving rise to the formation of a
$^{14}$N-pocket with  $^{14}$N abundances by  mass of about  0.45 (the
most abundant element in the pocket) and an underlying $^{13}$C-pocket
with a maximum $^{13}$C abundance  of 0.06.  The mass range over which
the  $^{14}$N-rich region extends  amounts only  to $\approx  2 \times
10^{-6}, M_{\sun}$. During  the interpulse period, the $^{13}$C-pocket
is radiatively burnt at  relatively low temperatures (about $80 \times
10^6$~K)  before  the  onset  of  the  next  pulse  via  the  reaction
$^{13}$C$(\alpha,n)^{16}$O,  the main neutron  source reaction  in AGB
stars.  The $^{14}$N-pocket is engulfed by the helium-flash convection
zone during the next thermal pulse and burnt via the $^{14}$N$(\alpha,
\gamma)^{18}$F($\beta^+,  \nu)^{18}$O$(\alpha,  \gamma)^{22}$Ne  chain 
that  converts all the  abundantly present  $^{14}$N to  $^{22}$Ne. In
agreement  with Herwig  (2000), our  calculations show  that diffusive
overshooting at the base of  the convective envelope is a process that
naturally leads  to the formation of $^{13}$C-  and $^{14}$N- pockets,
which is a fundamental issue regarding the formation of heavy elements
through the slow neutron capture process --- see, for instance, Lugaro
et al.  (2003) for a recent discussion.

\subsection{Evolution  during the  last thermal  pulse and  born again: 
the burnt of hydrogen envelope}

As  a result  of  the mass-loss  episodes,  the mass  of the  hydrogen
envelope  is reduced  to such  an  extent that  the thermally  pulsing
progenitor star leaves  the AGB and evolves into  the planetary nebula
regime at  large \teff\ values.   During this phase of  the evolution,
helium-shell  burning increases  gradually and  when  hydrogen burning
becomes virtually  extinct at the  beginning of the white
dwarf  cooling  branch,  the  post-AGB remnant  experiences  the  last
helium-shell  flash  that gives  rise  to  the short-lived  born-again
episode. The mass of hydrogen that it is left in the very outer layers
at the start of the last thermal pulse amounts to $7 \times 10^{-5} \,
M_{\sun}$.  It  is worth mentioning  that very few  detailed numerical
simulations through  this complicated regime exist  in the literature.
In  particular, evolutionary  calculations that  include  hydrogen and
helium burning combined with time-dependent mixing have been performed
initially by Iben \& MacDonald (1995) and by Herwig et al.  (1999) for
the  situation in  which diffusive  overshooting is  considered. Also,
Lawlor \& MacDonald  (2003) have recently presented a  grid of stellar
evolutionary  calculations  for  the  born-again phenomenon  aimed  to
explain the observational characteristics of born-again stars. In what
follows, we  report the main  predictions of our calculations  for the
born-again  phase,  particularly emphasizing  on  the  changes in  the
abundances of  the different chemical  species that take  place during
this brief evolutionary phase.

The Hertzsprung-Russell diagram focusing  onto the last helium thermal
pulse and the subsequent born-again evolutionary phase is displayed in
Fig. \ref{m2.7-hr-pfp.eps}.  Selected evolutionary stages are labelled
by  letters  along  the  evolutionary  track.   The  inset  shows  the
time-dependence of hydrogen-  (CNO-cycle reactions) and helium-burning
luminosities, again  selected evolutionary stages  are correspondingly
labeled in this curve.  Note  that remarkable changes in the structure
of the star  take place on extremely short  time-scales. For instance,
it takes 0.4~yr for  the star to develop hydrogen-burning luminosities
as high  as $10^8\, L_{\sun}$ (point C  in Fig. \ref{m2.7-hr-pfp.eps})
after protons begin to be engulfed by the outward-growing helium-flash
convection  zone (point  B).  Most  of the  hydrogen  envelope burning
occurs between points C and E in about 1 month.  Specifically,  
at point E, after 0.9 yr have
elapsed from the onset of the last helium thermal pulse (point A), the
mass  of the  residual hydrogen  envelope has  reduced to  $10^{-7} \,
M_{\sun}$.  Between points E and F, hydrogen burning becomes gradually
extinct and the mass of hydrogen remaining in the star amounts to $1.3
\times 10^{-8}  \, M_{\sun}$. Approximately 0.6 yr  later, the remnant
reaches  the point  of maximum  effective temperature  at  $\log T_{\rm
eff}= 4.8$ for the first time after the helium flash.  Afterwards, the
evolution proceeds into the red  giant domain somewhat more slowly. In
fact, the effective temperature decreases  to 10000 K over a period of
about 20 yr,  and to 5200K in about  50 yr and the radius  of the star
increases to 30.2 and $125 \, R_{\sun}$, respectively.

It  is worth noting  that our  born-again time-scale  of 20--40  yr is
larger  than the  evolutionary  time-scale of  the born-again  Sakurai
object (V4334~Sgr),  which has evolved from the  pre-white dwarf stage
into  a  AGB  giant  star  in  only  about  6  yr.   The  evolutionary
calculations of  Herwig (2001), which  are based on the  standard MLT,
also   predict  too  large   born-again  time-scales   (typically  350
yr). Herwig has found that the very short born-again evolutionary time
of V4334~Sgr  can be  reproduced by stellar  models if  the convective
mixing efficiency in the helium-flash  convection zone is reduced by a
factor  of  100  below  that  obtained from  the  MLT.   Although  our
evolution  time-scales are larger  than observed  ones, they  are much
shorter than those obtained by  Herwig (2001), but without invoking an
additional reduction  in the mixing efficiency.  We  remind the reader
that  in our  calculation  the double-diffusive  MLT  for fluids  with
composition  gradients (Grossman  \&  Taam 1996)  has  been used\footnote{
It is clear that the big difference in the evolutionary time scales between the both set of calculations
cannot be entirely attributed to the fact that the mass value of our model 
is not strictly the same as that
analyzed by Herwig 2001.}.    
Another  prediction  of   our  calculations   is  the
occurrence of  a double-loop in the  Hertzsprung-Russell diagram. That
is, the star reaches red giant  dimensions for a second time after the
onset of the last helium thermal pulse and before finally returning to
the  white dwarf  cooling track,  a  behaviour reported  by Lawlor  \&
MacDonald  (2003) and  Herwig (2003)  in  the case  of low  convective
mixing efficiency. In particular, this  second return to the AGB takes
about 350~yr in  our calculations.  In the light  of these results, we
judge that a more comprehensive comparison between the standard mixing
length theory  and the double-diffusive mixing  length theory deserves
to be done.  Such a comparison would, however, carry us too far afield
and, consequently, we postpone it to a forthcoming work.

\begin{figure*}
\centering
\includegraphics[clip,width=500pt]{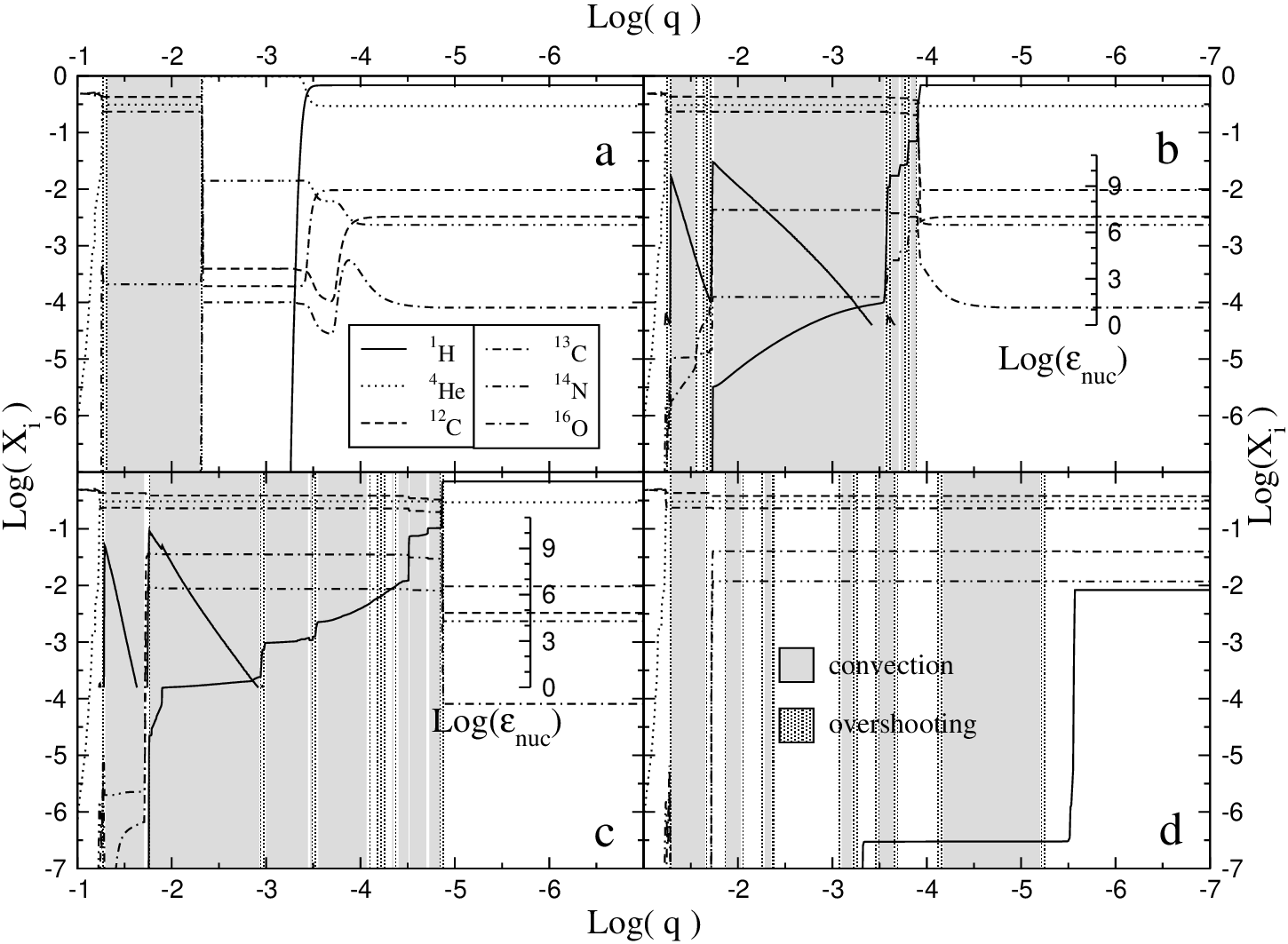}
\caption{Internal  abundance   distribution   of  $^{1}$H,   $^{4}$He, 
$^{12}$C, $^{13}$C, $^{14}$N  and $^{16}$O as a function  of the outer
mass  fraction $q$  for the  $0.5885 \,  M_{\sun}$ remnant  at various
selected epochs  during the course  of the last helium  thermal pulse.
Panel {\sl a} corresponds to  the moment just before the occurrence of
the  peak  helium-burning  luminosity  (point  A in  figure  4).   The
situation  when the outward-moving,  helium-flash convection  zone has
reached the base of the hydrogen-rich envelope (between points B and C
in figure 4)  is illustrated by panel {\sl b}.   Note that protons are
ingested into  deeper and  hotter layers. As  a result of  the ongoing
hydrogen burning a small radiative and salt-finger zone establishes at
log q $\approx$ -1.6, which  splits the convection zone into two.  The
situation two weeks later is visualized in panel {\sl c}.  During this
time,  the  convection zone  powered  by  hydrogen burning  propagates
outwards. Finally,  panel {\sl  d} illustrates the  abundance profiles
near  point F  in  figure 4.   Only  traces of  hydrogen remain  after
hydrogen burning.   Nuclear energy release due to  hydrogen and helium
burning are also shown in panel {\sl b} and {\sl c}.}
\label{4graf.eps}
\end{figure*}

During the last helium thermal  pulse, profound changes in the chemical
structure take  place as a  result of the vigorous  nuclear processing
and  mixing episodes.   Such changes  are crucial  for  the subsequent
evolution  of  the  star.   Because  of the  very  short  evolutionary
time-scales  characterizing  this phase,  a  numerical treatment  that
consistently  couples  the  equations  of  nuclear  changes  with  the
equations for  time-dependent mixing  processes, like the  one adopted
here, is indeed required for  a realistic description of the abundance
changes.  A  complete coverage of the inner  chemistry variations that
take  place  during  this  short  evolutionary stage  is  provided  in
Fig. \ref{4graf.eps}, which  shows the chemical abundance distribution
at four  selected evolutionary stages  during the last  thermal pulse.
Specifically, the  abundances by mass of  $^{1}$H, $^{4}$He, $^{12}$C,
$^{13}$C, $^{14}$N and $^{16}$O are  ploted in terms of the outer mass
fraction $q$. Grey  and shaded regions mark the  domains of convection
and overshooting.  Panel {\sl  a} shows the chemical stratification at
the  start of  helium thermal  pulse (point  A in  figure 4).   In the
outermost layers the chemical composition corresponds to that fixed by
dredge-up episodes during  the AGB phase.  In the  pure helium buffer,
the relatively large abundance  of $^{14}$N reflects the efficiency of
hydrogen burning during previous evolutionary phases in processing CNO
elements  into  $^{14}$N.   Because  of  the large  amount  of  energy
resulting from  helium burning, an  outward-growing  convective region
develops.   In about  1~yr, the  outer edge  of this convective region
reaches the base of the  hydrogen-rich envelope.  As a result, protons
begin to be transported  downwards into hotter and carbon-rich layers,
where they are captured  via the $^{12}$C($p, \gamma)^{13}$N reaction.
The ensuing vigorous hydrogen burning forces
the development of an entropy barrier. As a result, the original convective
region is split at $\log q \approx  -1.6$ in  two distinct  convective
shells. The intermediate region is radiative, but also presents a salt-
finger instability at its top. This  situation is illustrated  
by panel {\sl  b}.  Hydrogen
and helium burning  take place at the base  of such convective layers.
Note the  large amount of $^{13}$C  in the convection  zone powered by
hydrogen burning (upper convection zone).  In the hotter, helium-flash
convection  zone  $^{13}$C (and  $^{14}$N)  is  destroyed by  $\alpha$
captures and neutrons are released.
Note as well that, as previously emphasized, the use of a simultaneous
treatment  of  mixing and  burning  is  indeed  required during  these
stages, as  reflected by the hydrogen  profile.    

Panel  {\sl  c}  depicts  the   situation  two  weeks
later. During this  time, the outer edge of  the upper convection zone
propagates further outwards in mass, mixing downwards protons from the
original   (unprocessed)  hydrogen-rich   envelope.   Note   both  the
persistence  of  the  small  salt-finger  region  separating  the  two
convection  zones  and  the  increase  in the  $^{13}$C  and  $^{14}$N
abundances in  the upper convection  zone due to  hydrogen processing.
The surface and  upper convection zones merge temporarily  (point E in
Fig. \ref{m2.7-hr-pfp.eps}), causing  the convectively unstable region
to  extend from  the  hydrogen-burning zone  essentially  to the  very
surface  of the  star.  Only a very thin unprocessed hydrogen-rich 
envelope
remains, which is expected to be diluited by surface convection when the star
returns to the AGB. Finally,  panel  {\sl d}  shows the  abundance
profiles near  point F in Fig.  \ref{m2.7-hr-pfp.eps}.  Here, hydrogen
burning is  virtually extinct and the remaining  hydrogen mass amounts
to only  $1.3 \times 10^{-8} \,  M_{\sun}$.  Note also  that traces of
hydrogen are  present in layers even  as deep as $3  \times 10^{-4} \,
M_{\sun}$  below  the  stellar  surface.   Interestingly  enough,  the
content of $^{13}$C left in the  whole star is sizeable and amounts to
$4.5  \times 10^{-4}\,  M_{\sun}$, that  is 4--5  orders  of magnitude
larger than the hydrogen mass of the star.

\subsection{Evolution towards the PG1159 state} 

After  the last  thermal  pulse, as  the  star evolves  back to  giant
dimensions,  either  convective  dilution  or  mass  loss  during  the
quiescent helium-burning phase (Iben et al. 1983) are expected to erode
the tiny layer of  original envelope material, exposing the underlying
hydrogen-deficient  layers.  Thus,  after the  second loop  in  the HR
diagram, the remnant evolves to the  region of the PG1159 stars with a
surface chemical composition similar to that shown in panel {\sl d} of
Fig.   \ref{4graf.eps}. Specifically, in  the outer  layers, $^{4}$He,
$^{12}$C and $^{16}$O  are by far the dominant  species with abundance
by  mass  ($^{4}$He,$^{12}$C,$^{16}$O)= (0.306,0.376,0.228).   Amongst
the main  remaining constituents are $^{13}$C,  $^{14}$N and $^{22}$Ne
with mass fractions of 4,  1.2 and 2.1 \%, respectively.  $^{13}$C and
$^{14}$N are present from the outermost layers down to the base of the
former upper convection  zone at $0.01 \, M_{\sun}$  below the stellar
surface. At  deeper layers, these  chemical species have  been already
depleted via $\alpha$ captures.  The  large mass fraction of oxygen is
an indication  of the occurrence of diffusive  overshooting during the
thermally pulsing AGB phase (see also Herwig et al. 1999).  The final 
surface composition of
our models is also in line with surface abundance patterns observed in
most hot, hydrogen-deficient  post AGB stars such as  PG1159 stars and
central stars of  planetary nebulae of spectral type  WC (Koesterke \&
Hamann  1997; Dreizler \&  Heber 1998;  Werner 2001).   The remarkable
agreement between the $^{14}$N abundance predicted by our calculations
and the one  detected by Dreizler \& Heber (1998) in  five out of nine
PG1159 stars  strongly supports the hypothesis that  these stars would
be AGB descendants  that have experienced a born-again  episode. As we
have already seen, mixing and  burning of protons in the helium-flash
convective  region are  indeed required  to synthesize  $^{14}$N  in a
carbon- and oxygen-rich environment.

\begin{figure}
\centering
\includegraphics[clip,width=250pt]{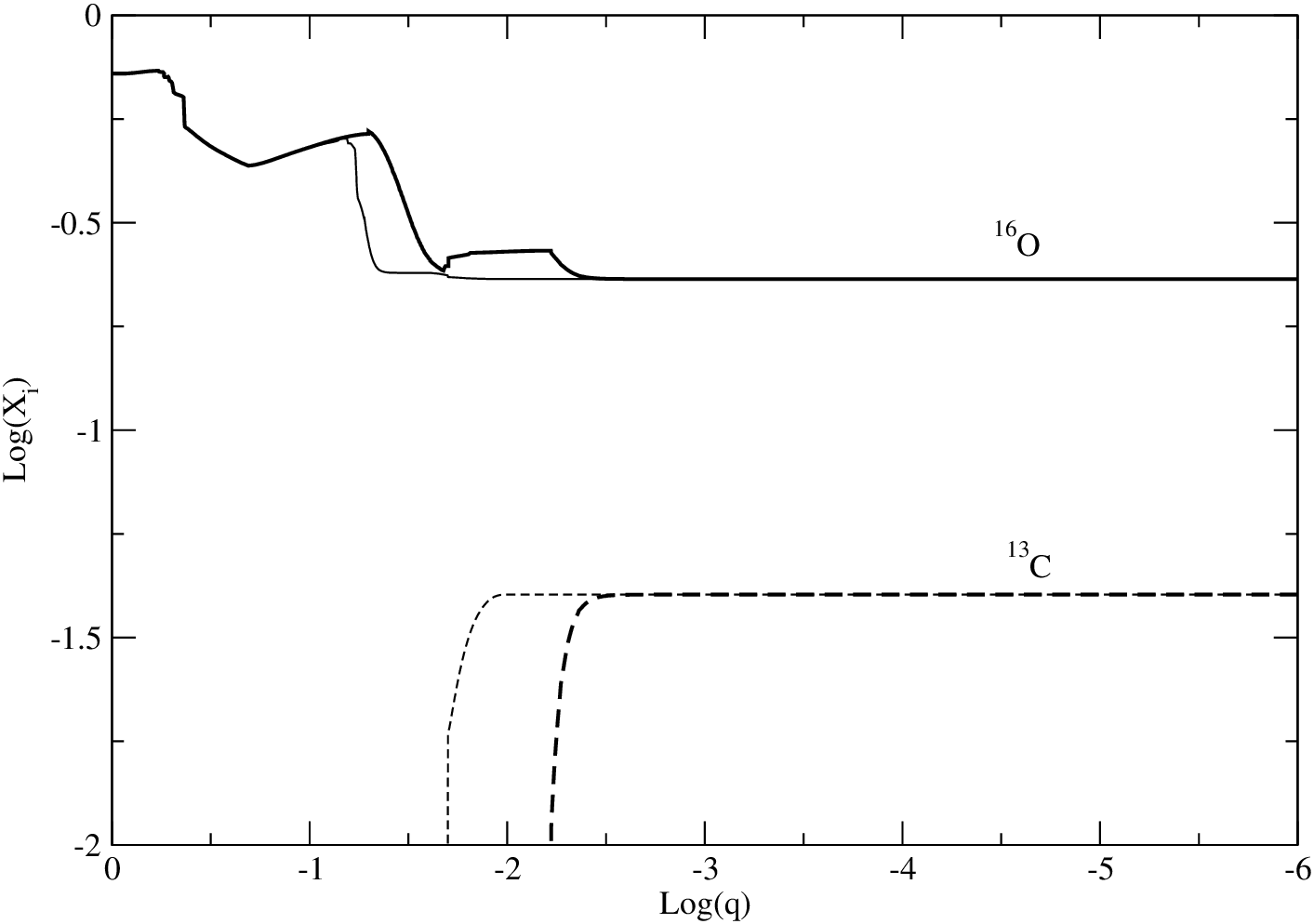}
\caption{Evolution of the internal abundances of $^{13}$C and $^{16}$O  
(dashed  and solid  lines respectively)  across the  PG1159  domain in
terms  of the outer  mass fraction  $q$ for  the $0.5885  \, M_{\sun}$
remnant. Thin and  thick lines correspond, respectively, to a model at
\teff= 71000~K  before  the point of maximun  effective temperature in
the  HR  diagram and  \teff=  125000~K  below  that point.   Note  the
formation of a bump in the oxygen distribution as a result of $\alpha$
captures by $^{13}$C.}
\label{bochin.eps}
\end{figure}

\begin{figure*}
\centering
\includegraphics[clip,width=500pt]{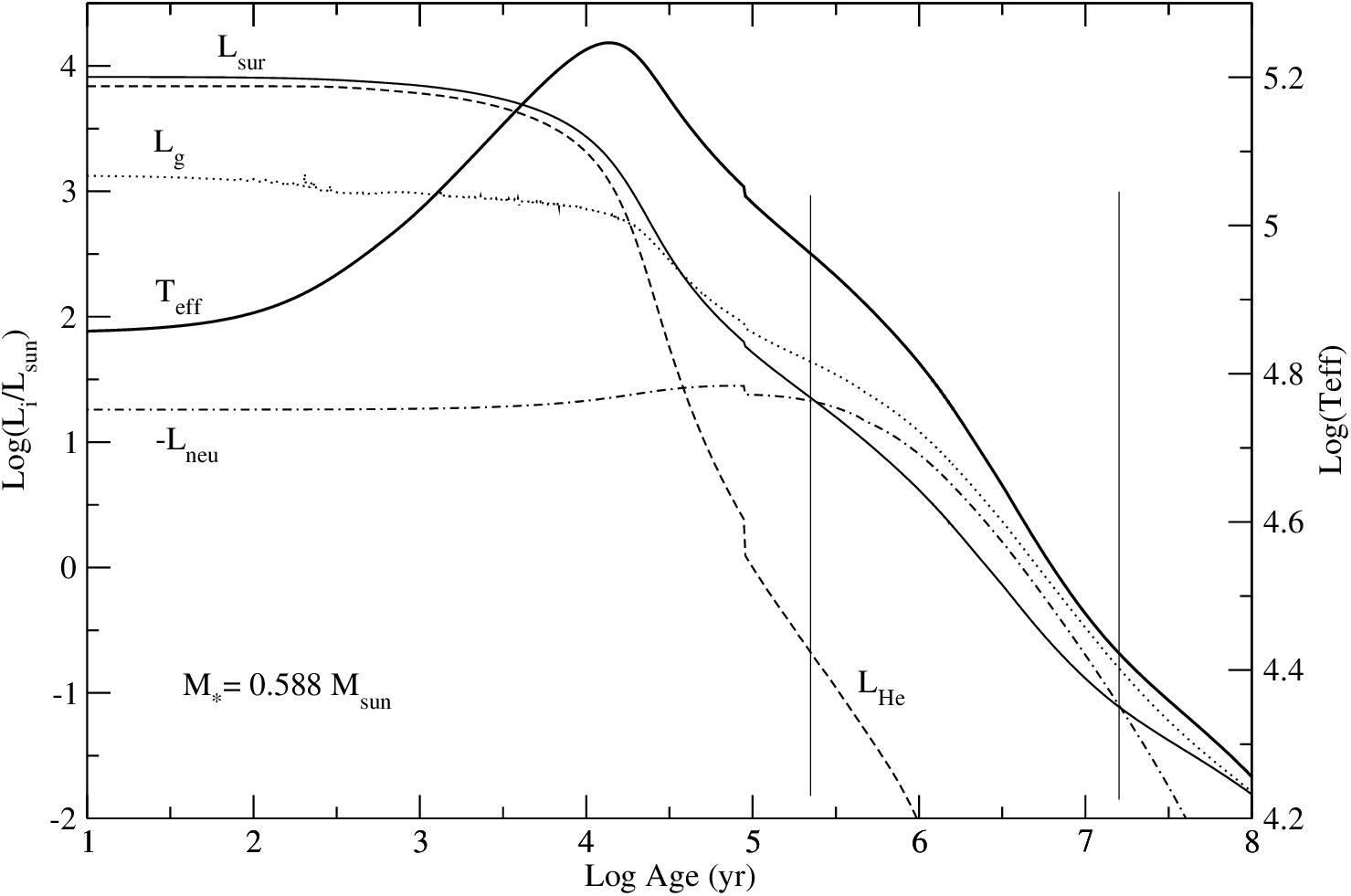}
\caption{Time-dependence  of  the  different  luminosity contributions 
(in solar units)  for the post born again  $0.5885\, M_{\sun}$ remnant
of  the evolution of  the $2.7\,  M_{\sun}$ star:  surface luminosity,
$L_{\rm  sur}$,  helium  burning  luminosity, $L_{\rm  He}$,  neutrino
losses,  $L_{\rm neu}$  and rate  of gravothermal  (compressional plus
thermal) energy  release $L_{\rm g}$.  The evolution  of the effective
temperature (right  scale) is also  plotted. Time is in  years counted
from  the moment  at which  the remnant  reaches \teff=  70000~K.  The
vertical lines bracket the  domain where neutrino losses exceed photon
luminosity.}
\label{lumis.eps}
\end{figure*}

Finally, we note that as  the evolution proceeds through the domain of
the PG1159 stars,  the changes in the chemical  composition take place
as a result  of nuclear burning.  Fig. \ref{bochin.eps}  is an example
of  this.  Note  in particular  that in  a region  stretching  from $3
\times 10^{-3}$ to $10^{-2} \,  M_{\sun}$ below the stellar surface, a
bump in  the oxygen distribution is  built up as a  result of $^{13}$C
burning  via  the  $^{13}$C$(\alpha,n)^{16}$O  reaction.   This  is  a
consequence  of the  fact that  before the  hydrogen-deficient remnant
reaches the point of maximum  effective temperature in the HR diagram,
the temperature  at the tail  of the $^{13}$C distribution  exceeds 85
$\times    10^{6}$~K,     which    is    high     enough    for    the
$^{13}$C$(\alpha,n)^{16}$O reaction to operate.   As a last remark, we
note that  during the  early PG1159 stage,  the helium content  in the
star is  reduced from $8.5 \times  10^{-3}$ to $5.2  \times 10^{-3} \,
M_{\sun}$ as a result of helium burning.  The last quoted value is the
amount of  helium with  which the star  enters the white  dwarf domain
after helium burning becomes extinct.  Because we have not invoked any
additional  mass loss  after  the born-again  episode,  the amount  of
helium left  in the  white dwarf ($5.2  \times 10^{-3} \,  M_{\sun}$ )
should  be  considered as  an  upper  limit  for the  particular  case
analyzed in  this work (see next  section for the  effect of mass-loss
episodes during the hot stages of post-AGB evolution).

In Fig. \ref{lumis.eps}, we show  as a function of time the luminosity
contributions due  to helium  burning $(L_{\rm He})$,  neutrino losses
$(L_{\rm neu})$, surface  luminosity $(L_{\rm sur})$, and gravothermal
energy  release $(L_{\rm  g})$ for  the hydrogen-deficient  $0.5885 \,
M_{\sun}$  remnant.  The  abscissa  covers  the  time  span  from  the
pre-white  dwarf state,  before the  maximum  \teff\ point  in the  HR
diagram,   to   beyond  the   domain   of   the   variable  DB   white
dwarfs. Luminosities are given in solar units; the age is counted from
the moment at which the remnant reached \teff= 70000~K.  Some features
are worthy of a comment. To begin with, note that at early times it is
mostly helium  burning that contributes  to the surface  luminosity of
the star.  This is  true for the  first 15000--20000~yr  of evolution,
shortly after the remnant reaches the maximum \teff\ value, and begins
to  decrease   as  the  star   approaches  the  white   dwarf  domain.
Afterwards, the  contribution of  helium burning declines  steeply and
the evolution of  the star is dictated essentially  by neutrino losses
and  the  release  of   gravothermal  energy.   At  the  \teff\  value
characterizing    PG1159-035   ($\log   T_{\rm    eff}\simeq   5.12$),
gravothermal energy  is the main energy  source of the  star. But, for
the  coolest pulsators  in  the GW~Vir  strip  neutrino losses  exceed
photon luminosity.   This is an  important feature since it  allows to
use cool pulsating PG1159 stars to constrain  neutrino emission
processes  from measurements  of the  rate  of period  change ---  see
O'Brien et al.  (1998) and  O'Brian \& Kawaler (2000) for a discussion
of  this exciting  issue.  Note  that neutrino  losses  constitute the
primary  cooling  mechanism  over   a  period  of  about  $1.5  \times
10^{7}$~yr,  during which  the  star evolves  through the temperature
range  90000 K $\gtrsim$  \teff  $\gtrsim$ 25000~K.   It  is  worth
mentioning at this point  that the lower temperature limit corresponds
to  the domain  of  the hot  pulsating  DB white  dwarfs. Thus,  these
variable stars should also  be potentially useful to place constraints
on plasmon  neutrino emissivity, as  recently emphasized by  Winget et
al. (2004).

\begin{figure}
\centering
\includegraphics[clip,width=250pt]{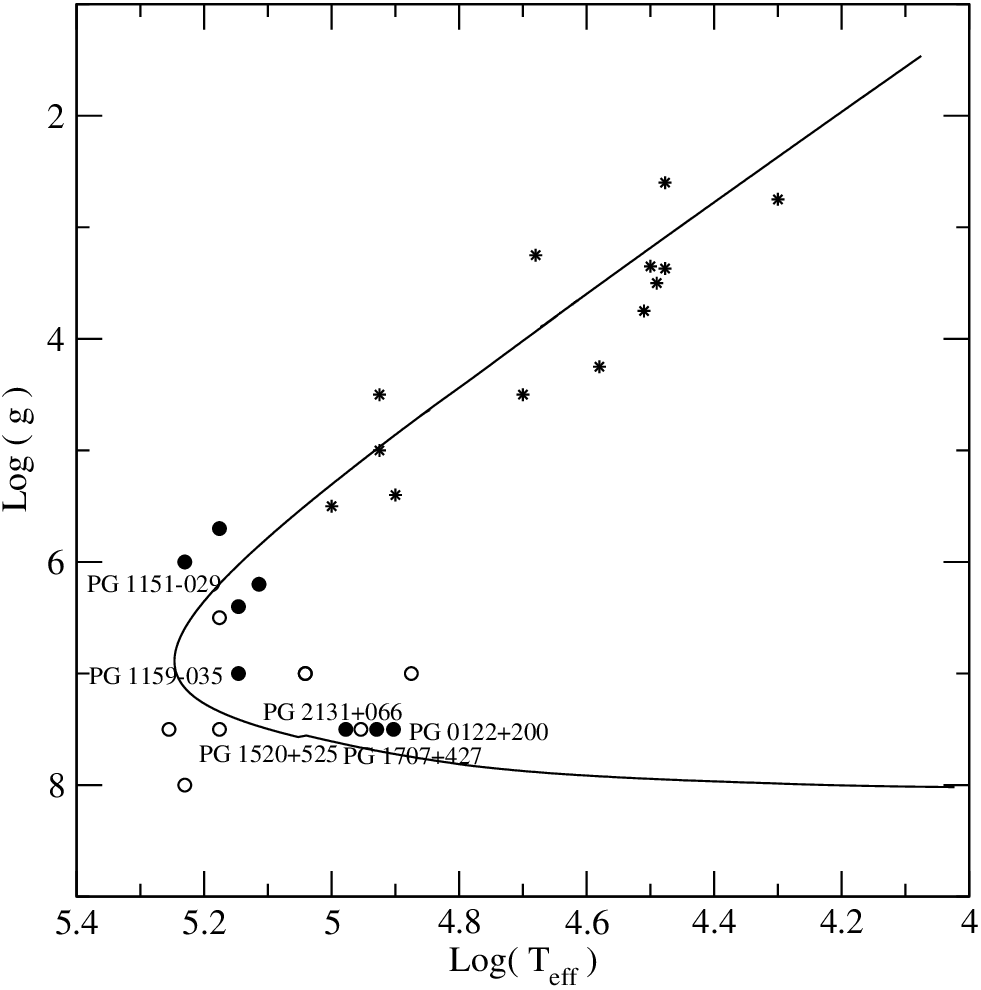}
\caption{Evolution of the surface gravity (in cgs units) as a function 
of  the  effective temperature  for  the  post  born-again $0.5885  \,
M_{\sun}$  remnant   and  its   comparison  with  some   observed  hot
hydrogen-deficient  stars.  Filled  (open)  circles  denote  pulsating
(non-pulsating)  PG1159 stars,  and  asterisks stand  for observed  WC
stars (Werner et al. 1997; Werner 2001).}
\label{grave.eps}
\end{figure} 

Finally,    in   Fig.    \ref{grave.eps}    we   show    the   surface
gravity--effective  temperature   diagram  for  our   post  born-again
$0.5885\,  M_{\sun}$  sequence together  with  observational data  for
hydrogen-deficient PG1159  and WC stars,  as taken from Werner  et al.
(1997) and Werner (2001).   In particular, the pulsating nitrogen-rich
PG1159  stars  analyzed  by  Dreizler  \&  Heber  (1998)  ---  namely,
PG1159-035, PG2131+066,  PG1707+427 and PG0122+200  --- are consistent
with a stellar  mass somewhat lower than that  characterizing our post
born-again sequence.

\subsection{White dwarf evolution} 

\begin{figure*}
\centering
\includegraphics[clip,width=500pt]{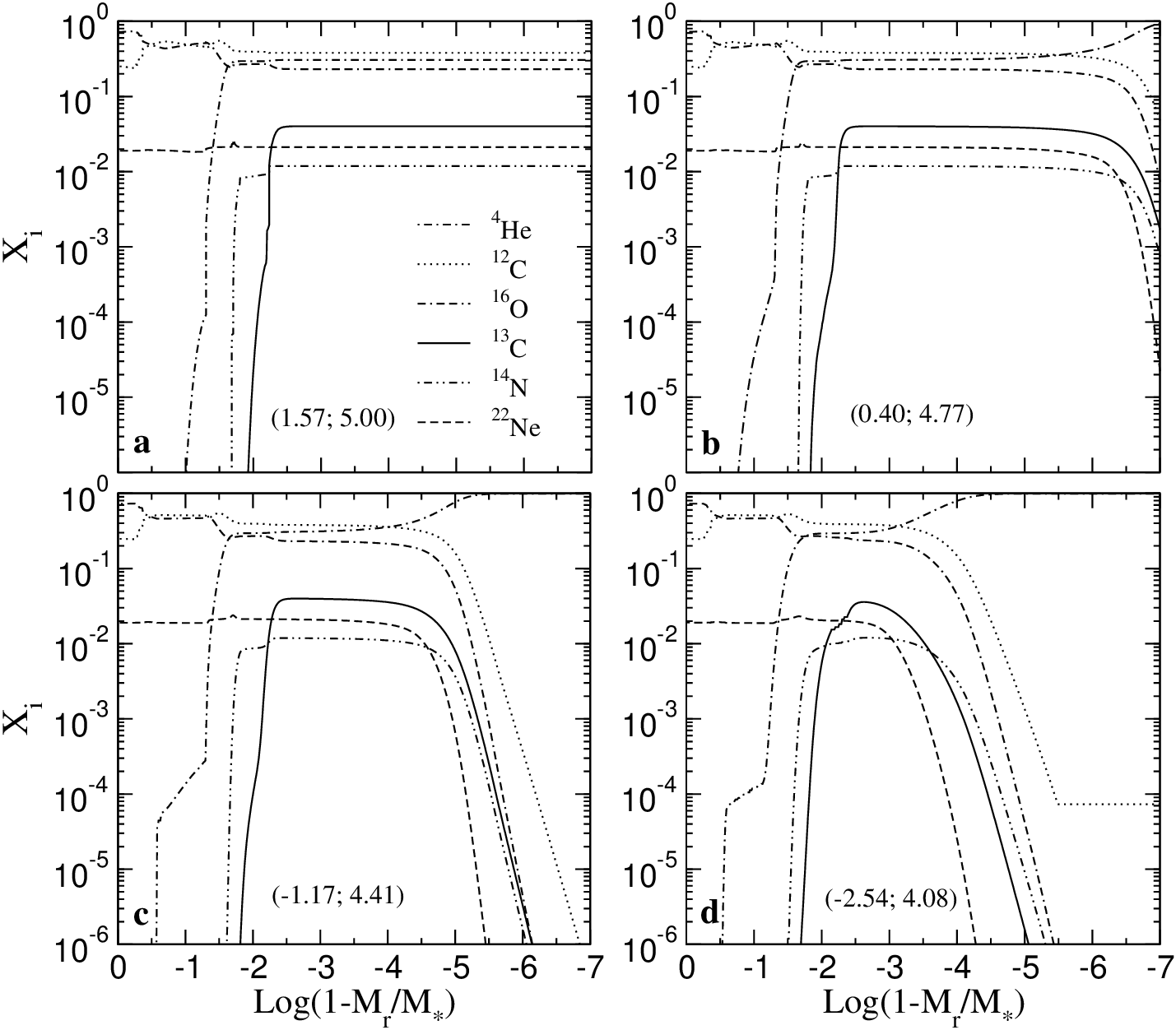}
\caption{Abundance  by mass  of  $^4$He, $^{12}$C, $^{16}$O, $^{13}$C, 
$^{14}$N and  $^{22}$Ne in  terms of the  outer mass fraction  for the
$0.5885 \, M_{\sun}$ remnant at four selected white-dwarf evolutionary
stages, characterized by values of $\log L/L_{\sun}$  and $\log T_{\rm
eff}$ (the corresponding values are given in parentheses).  Panel {\sl
a} corresponds to the start of the cooling branch, panel {\sl c} to an
evolutionary stage during  the DB instability strip and  panel {\sl d}
to the domain  of the carbon-enriched DQ white  dwarfs.  Panel {\sl d}
illustrates the significant carbon enrichment in the surface layers as
a result of  convective dredge-up of the carbon  diffusive tail by the
superficial helium convection zone. It is clear that element diffusion
strongly   modifies   the    internal   chemical   profiles   of   the
hydrogen-deficient white dwarf.}
\label{diffusion-log.eps}
\end{figure*}

Once  helium  burning becomes  virtually  extinct  at \teff  $\approx$
100000~K, after  100000~yr of evolution  from the last  helium thermal
pulse,  the  hydrogen-deficient  remnant  settles  upon  its  terminal
cooling  track.  Here,  its  chemical abundance  distribution will  be
strongly modified by the various diffusion processes acting during the
white     dwarf     evolution.      This     can    be     seen     in
Fig. \ref{diffusion-log.eps} which  illustrates the mass abundances of
$^4$He, $^{12}$C, $^{16}$O, $^{13}$C,  $^{14}$N, and $^{22}$Ne for the
$0.5885  \, M_{\sun}$  white dwarf  as a  function of  the  outer mass
fraction  at  various  epochs  characterized  by  values  of  $\log{L/
L_{\sun}}$ and $\log{T_{\rm eff}}$ (the corresponding values are given
in parentheses for  each one of the panels).  Panel  {\sl a} shows the
chemical stratification  at the start  of the cooling track  after the
star has reached the point of maximum \teff\ at high luminosities.  In
the outermost layers  the chemical composition corresponds essentially
to that  emerging from the mixing  and burning events  during the last
helium  thermal  pulse and  subsequent  born-again episode.   Rapidly,
gravitational  settling causes  helium  to float  to  the surface  and
heavier elements  to sink.  In  fact, in about $1.5  \times 10^{6}$~yr
the star develops  a pure helium envelope of  nearly $6 \times 10^{-8}
\, M_{\sun}$ (panel  {\sl b}). By the time the  domain of the variable
DBs  is  reached  (after   $1.77  \times  10^{7}$~yr  of  white  dwarf
evolution; panel {\sl c}) gravitationally induced diffusion has led to
the development  of a double-layered  chemical structure characterized
by a pure helium envelope of $ 2.3 \times 10^{-6} \, M_{\sun}$ atop an
intermediate  remnant shell  rich in  helium, carbon  and  oxygen, the
relics of the last helium thermal pulse (see previous section).  About
$3.7 \times  10^{8}$ yr later, the  white dwarf reaches  the domain of
the   helium-rich,   carbon-contaminated   DQ   white   dwarfs.    The
corresponding chemical stratification is shown in panel {\sl d}.  Even
at  such   an  advanced  stage,   the  star  is  characterized   by  a
double-layered  structure.  In  particular, the  pure  helium envelope
amounts  to $1.9 \times  10^{-5} \,  M_{\sun}$.  Note  the significant
carbon  enrichment in  the surface  layers as  a result  of convective
dredge-up  of the  carbon  diffusive tail  by  the superficial  helium
convection  zone (see  later  in this  section).   The neutron  excess
characterizing  $^{13}$C (and also  $^{22}$Ne) partially  explains the
fact that  this element appreciably  diffuses downwards.  It  is clear
from these  figures that  diffusion processes substantially  alter the
chemical  abundance   distribution  in  the  course   of  white  dwarf
evolution.

\begin{figure}
\centering
\includegraphics[clip,width=250pt]{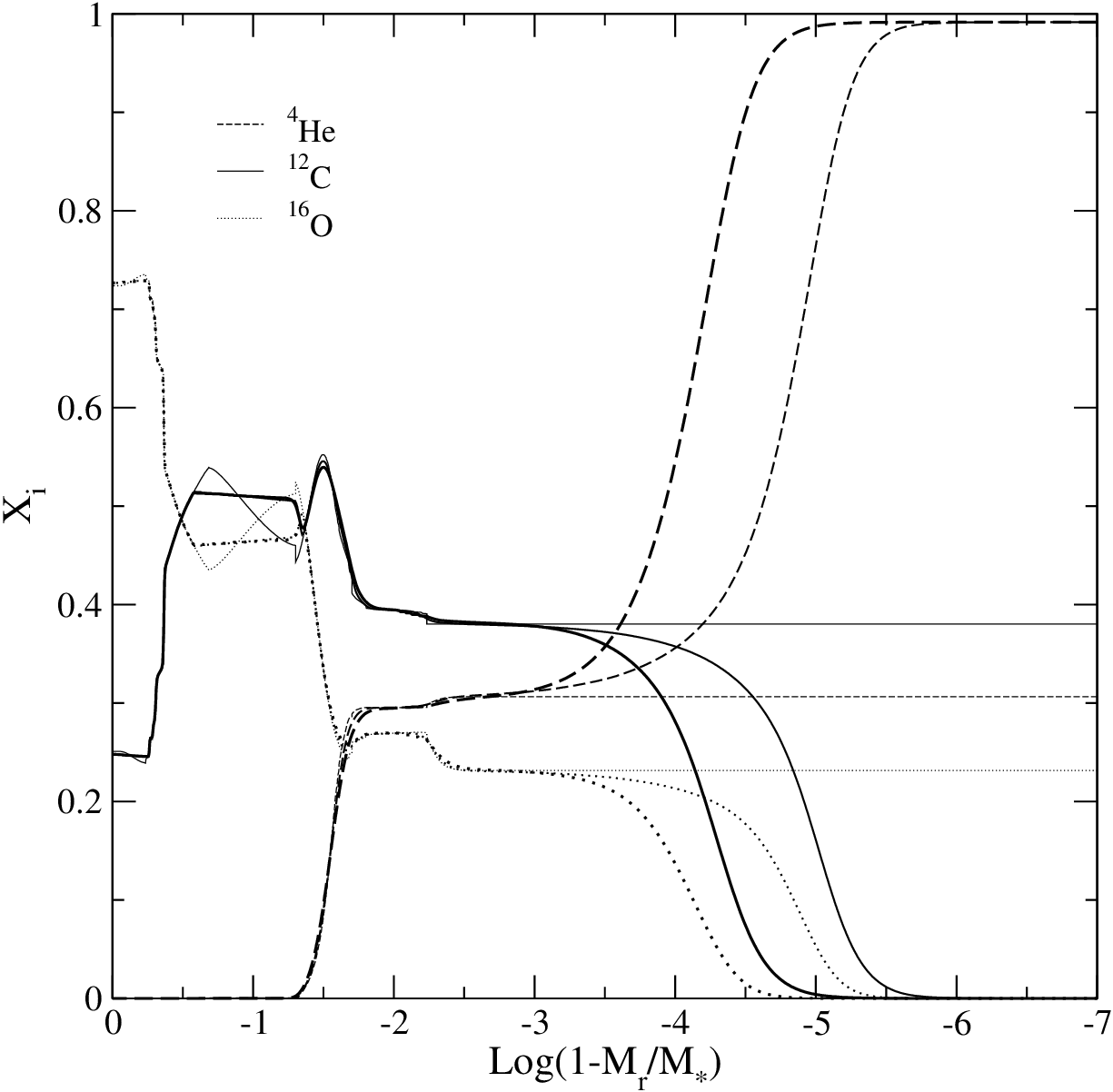}
\caption{Abundances by mass of  $^{4}$He, $^{12}$C and
$^{16}$O as a  function of the outer mass fraction  for the $0.5885 \,
M_{\sun}$ white-dwarf remnant having  a final helium content  of $5.2
\times  10^{-3} \,  M_{\sun}$. Initial  profiles at  the start  of the
cooling  branch (\lteff=  5.0)  are plotted  with  thin lines.   Later
stages,  denoted with normal  and heavy  lines, correspond  to \lteff=
4.41 and 4.28 respectively.}
\label{diffusion-lin.eps}
\end{figure}

The effect of element diffusion  on the main chemical constituents can
best be visualized in Fig. \ref{diffusion-lin.eps}, particularly the
formation  of   the  double-layered  chemical   structure.   The  mass
abundance of $^{4}$He,  $^{12}$C and $^{16}$O are shown  as a function
of the  outer mass fraction for  the $0.5885 \,  M_{\sun}$ white dwarf
remnant at  three evolutionary stages.  The  chemical profiles shortly
after  the remnant  reaches  the point  of  maximum \teff\  in the  HR
diagram  are  represented with  thin  lines  having various  patterns.
Later  evolutionary stages  around the  domain of  the  DB instability
strip (\lteff=  4.41 and 4.28)  are represented with normal  and heavy
lines.  We  stress again that the  helium content that is  left in the
star  at the start  of the  cooling branch,  after helium  burning has
virtually ceased, amounts to 5.2 $\times 10^{-3} \, M_{\sun}$. Because
we have not  invoked any additional mass loss  during the hot post-AGB
stages or  early during the cooling  branch, the quoted  value for the
final  helium mass  should be  considered as  an upper  limit  for the
particular  case of  evolution analyzed  here.   The diffusion-induced
double-layered structure at the domain  of the pulsating DBs is easily
recognizable.  Another feature worthy of comment is the mixing episode
that takes place in the  region below the intershell zone around $\log
q= -1$. In  fact, this region is characterized  by a inward-decreasing
mean molecular weight induced by the occurrence of overshooting during
the AGB thermally pulsing  phase.  The resulting salt-finger mixing is
responsible  for the redistribution  of the  chemical species  in that
region, as it is apparent from Fig.  \ref{diffusion-lin.eps}.

\begin{figure}
\centering
\includegraphics[clip,width=250pt]{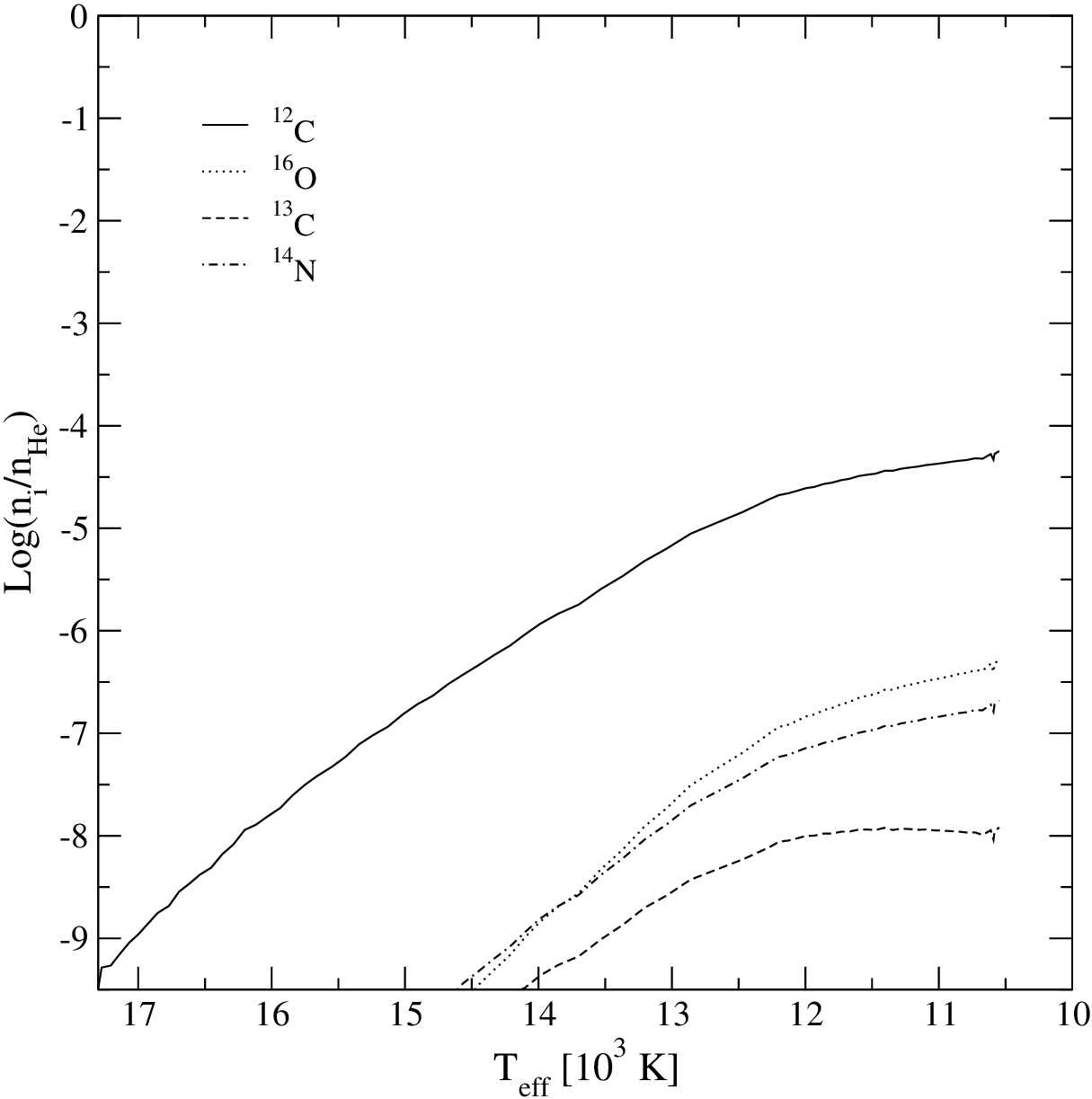}
\caption{Logarithm of the number density of surface $^{12}$C, $^{16}$O, 
$^{13}$C and  $^{14}$N relative to that  of $^{4}$He as  a function of
the effective temperature.}
\label{ninhe.eps}
\end{figure}

Finally, we  have extended the scope of  our evolutionary calculations
down  to the  domain  of  the carbon-enriched  DQ  white dwarfs,  thus
covering the  possible evolutionary connection  PG1159-DB-DQ. DQ white
dwarfs have  \teff\ values below  13000~K (Weidemann \&  Koester 1995)
and are  characterized by the presence  of trace amounts  of carbon in
their atmospheres with abundances  by number relative to helium, $\log
(n_{\rm C}/n_{\rm  He}$), ranging from $-7.3$ to  $-1.5$ (MacDonald et
al.  1998).  The presence of  traces of carbon observed in these stars
is widely believed  to result from convective dredge-up  of the carbon
diffusive tail by the superficial helium convection zone (Pelletier et
al. 1986; Koester et al. 1982).  By the time the DQ domain is reached,
our  white  dwarf  models  have developed  a  double-layered  chemical
structure  with  a pure  helium  mantle of  $  1.9  \times 10^{-5}  \,
M_{\sun}$, which  is almost  an order of  magnitude as massive  as the
helium mantle  characterizing our  models at the  beginning of  the DB
instability  strip.  However,  it is  less massive  than  required for
heavy  elements to  be abundantly  dredged-up  to the  surface by  the
inwards-growing superficial convection zone. This fact is reflected by
Fig.   \ref{ninhe.eps} which  shows  the surface  abundance by  number
relative to $^{4}$He of $^{12}$C, $^{16}$O, $^{13}$C and $^{14}$N as a
function of the effective temperature.  As the white dwarf cools down,
the base of  the convection zone moves deeper into  the star, with the
consequent further enrichment with heavy elements of the outer layers.
Note that  $^{12}$C is  by far the  most abundant  dredged-up element,
with abundances  far exceeding the  low carbon abundances  observed in
many DQ (see  next section).  Note also that  our calculations predict
the presence  of $^{13}$C and $^{14}$N, abundantly  created during the
last helium thermal  pulse, as well as $^{16}$O  in the atmospheres of
DQ white  dwarfs.  Finally,  $^{22}$Ne has diffused  so deep  into the
star (see \ref{diffusion-log.eps}, panel  {\sl d}) that it is expected
not to  be dredged-up to the surface  of these stars.

For  the sake of
completeness, we provide  in Table 1 some relevant  quantities for our
post born-again $0.5885 \,  M_{\sun}$ sequence.  Specifically, we list
from left  to right the  effective temperature, the  photon luminosity
(in solar units), the age (in  years) counted from the moment at which
the remnant  reaches \teff= 10000~K (after the occurrence of the double
loop at high luminosities),  the stellar radius (in  cm), the
surface  gravity,  and the  helium-burning  and neutrino  luminosities
(both in solar units). The  tabulation covers the stages following the
end of the born-again episode to  the domain of the DQ white dwarfs at
low \teff\ values.

\begin{center}
\begin{table*}
\caption{Selected stages for our post-born again $0.5885\, M{\sun}$ sequence.}
\centerline{}
\begin{tabular*}{140mm}{ccccccc}
\hline
\hline
$\log T_{\rm eff}$ [K] & $\log \left(\frac{L}{L_{\sun}}\right)$ & $\log{\tau}$ [yr] & $\log R_*$ [cm] & $\log g$ [cm s$^{-2}$]& $\log \left(\frac{L_{\rm He}}{L_{\sun}}\right)$ & $\log \left(-\frac{L_{\nu}}{L_{\sun}}\right)$\\
\hline
        4.0466 &         3.9940 &         0.3629 &        12.2700 &         1.3524 &         3.9908 &         1.2555\\
        4.1044 &         3.9928 &         0.7537 &        12.1539 &         1.5848 &         3.9884 &         1.2555\\
        4.1493 &         3.9916 &         0.9283 &        12.0634 &         1.7657 &         3.9865 &         1.2555\\
        4.2061 &         3.9898 &         1.0999 &        11.9489 &         1.9948 &         3.9836 &         1.2554\\
        4.2586 &         3.9878 &         1.2323 &        11.8429 &         2.2067 &         3.9804 &         1.2554\\
        4.3060 &         3.9856 &         1.3402 &        11.7470 &         2.3985 &         3.9770 &         1.2554\\
        4.3586 &         3.9826 &         1.4526 &        11.6402 &         2.6120 &         3.9724 &         1.2554\\
        4.4053 &         3.9795 &         1.5480 &        11.5453 &         2.8019 &         3.9675 &         1.2553\\
        4.4552 &         3.9756 &         1.6473 &        11.4435 &         3.0056 &         3.9611 &         1.2553\\
        4.5000 &         3.9714 &         1.7348 &        11.3519 &         3.1888 &         3.9542 &         1.2549\\
        4.5537 &         3.9655 &         1.8392 &        11.2416 &         3.4092 &         3.9442 &         1.2548\\
        4.6017 &         3.9593 &         1.9345 &        11.1425 &         3.6075 &         3.9328 &         1.2548\\
        4.6500 &         3.9519 &         2.0355 &        11.0421 &         3.8083 &         3.9200 &         1.2547\\
        4.7014 &         3.9424 &         2.1563 &        10.9346 &         4.0233 &         3.9003 &         1.2548\\
        4.7515 &         3.9315 &         2.3064 &        10.8290 &         4.2345 &         3.8734 &         1.2549\\
        4.8010 &         3.9221 &         2.5381 &        10.7251 &         4.4422 &         3.8384 &         1.2559\\
        4.8508 &         3.9156 &         2.7410 &        10.6225 &         4.6476 &         3.8360 &         1.2585\\
        4.9003 &         3.9007 &         2.8811 &        10.5159 &         4.8607 &         3.8366 &         1.2603\\
        4.9500 &         3.8804 &         2.9975 &        10.4064 &         5.0797 &         3.8240 &         1.2621\\
        5.0002 &         3.8549 &         3.1382 &        10.2933 &         5.3060 &         3.7937 &         1.2649\\
        5.0503 &         3.8224 &         3.2890 &        10.1768 &         5.5390 &         3.7610 &         1.2693\\
        5.1005 &         3.7774 &         3.4503 &        10.0538 &         5.7848 &         3.7121 &         1.2758\\
        5.1502 &         3.7134 &         3.6242 &         9.9224 &         6.0478 &         3.6410 &         1.2868\\
        5.2001 &         3.6081 &         3.8226 &         9.7700 &         6.3525 &         3.5196 &         1.3038\\
        5.2466 &         3.2645 &         4.1517 &         9.5053 &         6.8819 &         3.0875 &         1.3515\\
        5.2001 &         2.6917 &         4.4184 &         9.3118 &         7.2688 &         2.1510 &         1.4052\\
        5.1507 &         2.3614 &         4.5701 &         9.2454 &         7.4018 &         1.5060 &         1.4294\\
        5.1001 &         2.0635 &         4.7480 &         9.1978 &         7.4970 &         0.9067 &         1.4455\\
        5.0525 &         1.8007 &         4.9492 &         9.1615 &         7.5694 &         0.3861 &         1.4480\\
        5.0018 &         1.5620 &         5.1430 &         9.1435 &         7.6055 &        -0.2817 &         1.3667\\
        4.9512 &         1.2973 &         5.4055 &         9.1125 &         7.6675 &        -0.7823 &         1.3114\\
        4.9011 &         1.0430 &         5.6464 &         9.0855 &         7.7215 &        -1.2399 &         1.1936\\
        4.8517 &         0.7988 &         5.8578 &         9.0621 &         7.7684 &        -1.6699 &         1.0408\\
        4.8003 &         0.5493 &         6.0509 &         9.0403 &         7.8120 &        -2.2074 &         0.8467\\
        4.7502 &         0.3148 &         6.2117 &         9.0232 &         7.8461 &        -2.9096 &         0.6406\\
        4.7011 &         0.0919 &         6.3536 &         9.0100 &         7.8726 &        -3.7778 &         0.4345\\
        4.6511 &        -0.1312 &         6.4955 &         8.9984 &         7.8957 &        -4.8588 &         0.2115\\
        4.6015 &        -0.3477 &         6.6278 &         8.9893 &         7.9139 &        -6.0306 &        -0.0137\\
        4.5540 &        -0.5528 &         6.7585 &         8.9817 &         7.9291 &        -7.3201 &        -0.2475\\
        4.5045 &        -0.7646 &         6.9069 &         8.9748 &         7.9429 &        -8.9760 &        -0.5207\\
        4.4558 &        -0.9713 &         7.0723 &         8.9689 &         7.9547 &       -99.0000 &        -0.8388\\
        4.4035 &        -1.1917 &         7.2858 &         8.9633 &         7.9660 &       -99.0000 &        -1.2737\\
        4.3540 &        -1.3994 &         7.5261 &         8.9584 &         7.9757 &       -99.0000 &        -1.8102\\
        4.3030 &        -1.6132 &         7.7839 &         8.9536 &         7.9853 &       -99.0000 &        -2.4829\\
        4.2505 &        -1.8326 &         8.0244 &         8.9489 &         7.9948 &       -99.0000 &        -3.2861\\
        4.2024 &        -2.0322 &         8.2128 &         8.9454 &         8.0018 &       -99.0000 &        -4.1214\\
        4.1527 &        -2.2370 &         8.3788 &         8.9423 &         8.0080 &       -99.0000 &        -4.9905\\
        4.1027 &        -2.4419 &         8.5257 &         8.9398 &         8.0128 &       -99.0000 &        -5.8260\\
        4.0500 &        -2.6564 &         8.6674 &         8.9380 &         8.0166 &       -99.0000 &        -7.9835\\
        4.0232 &        -2.7651 &         8.7349 &         8.9373 &         8.0180 &       -99.0000 &       -99.0000\\
\hline
\end{tabular*}
\end{table*}
\end{center}

\section{Discussion}

The calculations  presented in this work cover  the whole evolutionary
stages involved in the formation of DB white dwarfs via the born-again
scenario starting  on the ZAMS.  In particular,  our calculations have
followed the model  star to very advanced stages  of evolution down to
the  domain of  the helium-rich  carbon-contaminated DQ,  the supposed
cooler descendants of DBs.  Hence,  the study opens the possibility of
assessing  the  evolutionary   connection  PG1159-DB-DQ  (Fontaine  \&
Brassard   2002)  in   the  framework   of   complete  evolutionary
calculations  that  take into  account  a  detailed  treatment of  the
physical processes  that lead  to the formation  of hydrogen-deficient
white dwarfs.  In this sense, the results presented here reinforce the
conclusions arrived at  in Fontaine \& Brassard (2002)  and Althaus \&
C\'orsico  (2004)   about  the  presence   of  a  diffusively-evolving
double-layered chemical  structure in pulsating DB  white dwarfs.  Our
results show that element diffusion proceeds very efficiently in these
stars, causing the  thickness of the pure helium  envelope to increase
by almost an order of magnitude  by the time the DQ domain is reached,
as compared with the situation  at the beginning of the DB instability
strip.

A novel  aspect of our  work is the  fact that the calculation  of the
evolutionary stages  prior to the  white-dwarf formation allows  us to
make sound  predictions of the  surface abundances expected in  the DQ
stars. In  this connection, we find  that $^{12}$C is by  far the most
abundant element that is  convectively dredged-up to the outer layers,
but with  abundances far exceeding the low  carbon abundances detected
in many DQs  (MacDonald et al. 1998)  by a big margin. If  we push our
results  to their  limits (see  below),  then the  plausibility of  an
evolutionary link between PG1159 and  DQ stars with {\sl low} detected
carbon  abundance appears  to be  unclear if  convective  dredge-up is
responsible for the  observed carbon in DQs.  Last  but not least, our
study  also  predicts  the  presence  of trace  amounts  of  $^{16}$O,
$^{13}$C and $^{14}$N in the atmospheres of these white dwarfs.

\begin{figure}
\centering
\includegraphics[clip,width=250pt]{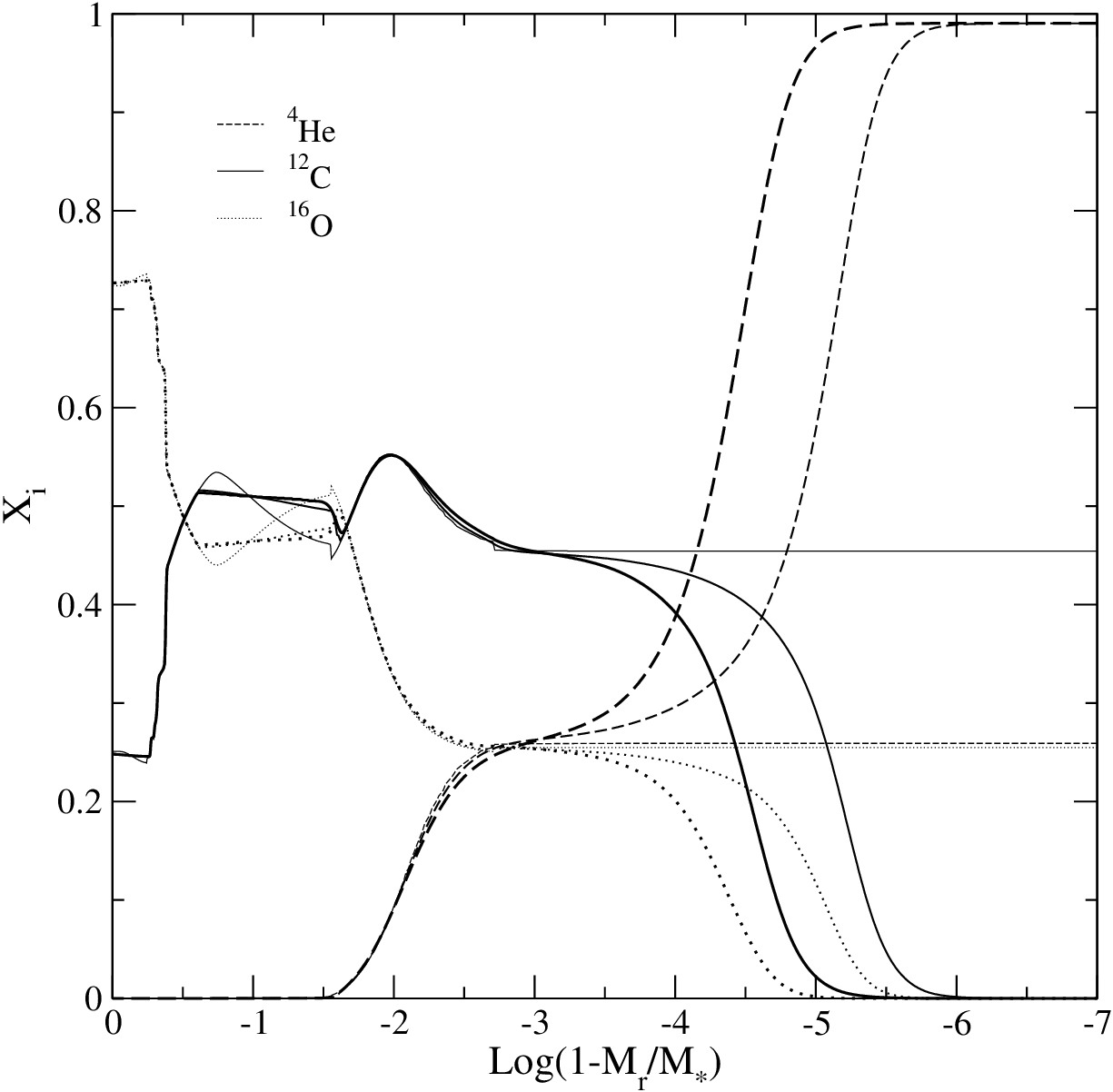}
\caption{Same as Fig.  \ref{diffusion-lin.eps}  but for a final helium
content of $1.4  \times 10^{-3} \, M_{\sun}$. The  initial profiles at
the start  of the cooling track  (\lteff= 4.89) are  plotted with thin
lines. Later  stages, denoted with normal and  heavy lines, correspond
to \lteff= 4.407 and 4.29 respectively.}
\label{diffusion-lin_2.eps}
\end{figure}

\begin{figure}
\centering
\includegraphics[clip,width=250pt]{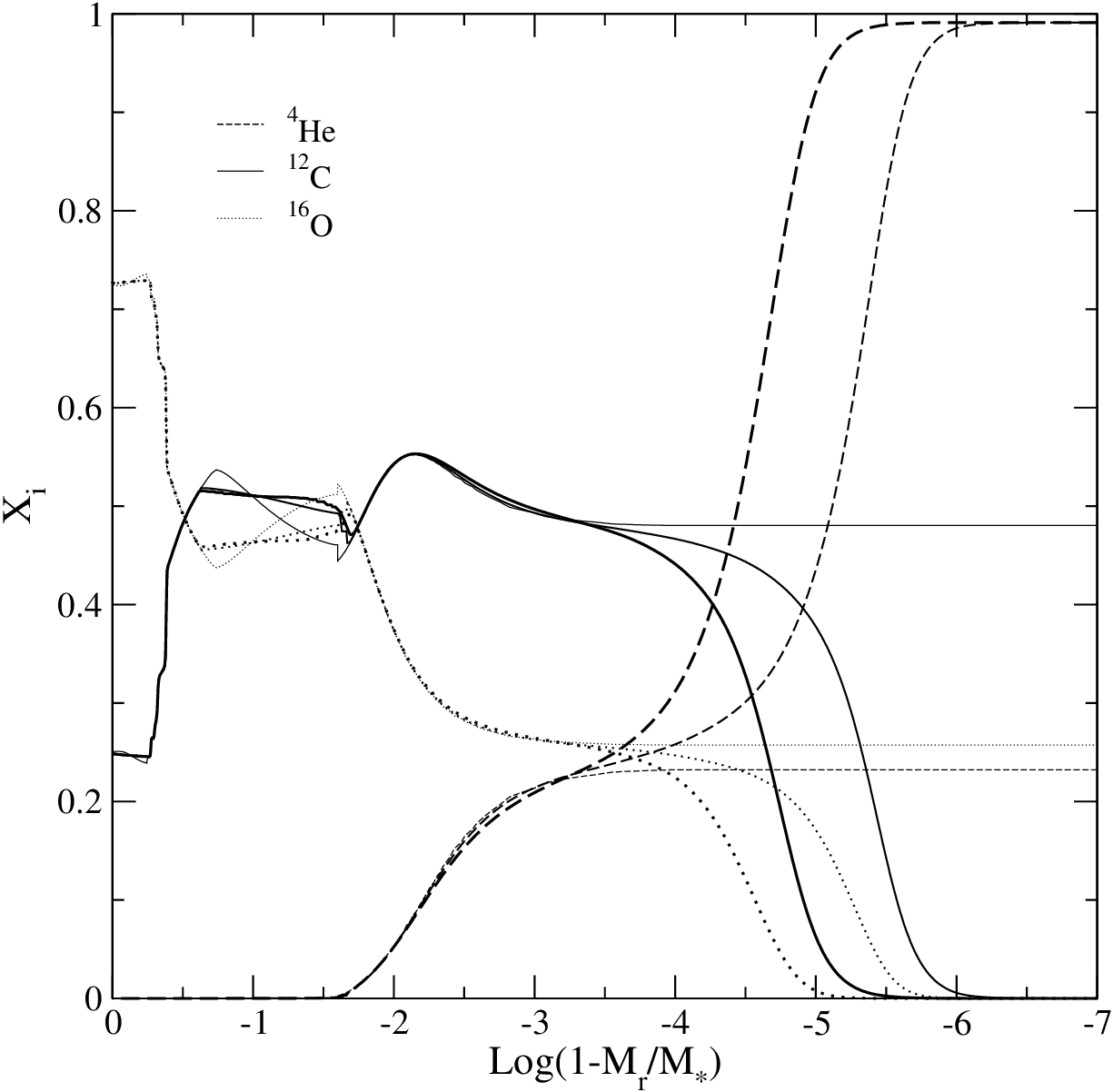}
\caption{Same  as Fig.  \ref{diffusion-lin.eps} but for a final helium
content of $9.4  \times 10^{-4} \, M_{\sun}$. The  initial profiles at
the start of  the cooling branch (\lteff= 4.93)  are plotted with thin
lines. Later  stages, denoted with normal and  heavy lines, correspond
to \lteff= 4.42 and 4.31 respectively. }
\label{diffusion-lin_3.eps}
\end{figure}

However, it  is conceivable that the occurrence  of mass-loss episodes
before and during the hot PG1159 stage could alter some aspects of the
above  mentioned   results  considerably.   In   this  regard,  recent
observational  evidence   hints  at  the   possibility  that  post-AGB
mass-loss episodes could markedly reduce 
the  mass of  the helium-rich  envelope.  Specifically,  the immediate
PG1159  predecessors,  the [WC]  stars,  which  roughly range  between
\teff= 30000 and  140000~K in the high-luminosity region  of the early
post-AGB  evolution  are  known  to  have  mass-loss  rates  of  about
$10^{-5.5}$ to $10^{-6.5} \, M_{\sun}/$yr that generally decrease with
increasing  \teff.   In  addition,  numerous hot,  low-gravity  PG1159
stars,  have  been reported  (Koesterke  \&  Werner  1998) to  exhibit
mass-loss  rates ranging  from $10^{-7}$  to  $10^{-8}\, M_{\sun}/$yr.
Finally, tentative evidence for  the persistence of mass-loss rates of
the same order  along the hot end of the  white dwarf cooling sequence
and down to  the domain of hot helium-rich white  dwarfs has also been
presented (Werner 2001).  Notably,  the existence of PG1159 stars with
a helium  content as low  as $1 \times  10^{-3} \, M_{\sun}$  has been
suggested by  asteroseismology in at least  one of these  stars with a
stellar  mass of  $0.6 \,  M_{\sun}$ (Kawaler  \& Bradley  1994), thus
implying the occurrence of  mass-loss during the evolution towards the
PG1159 phase.  These mass-loss rates  are indeed large enough to leave
in principle signatures in the further evolution of these stars.

To assess the  possible implications of such mass-loss  events for the
white  dwarf   evolution,  we   have  extended  our   calculations  by
considering two  further evolutionary sequences in which  mass loss is
addressed during the  hot post-AGB evolutionary stages.  Specifically,
we invoke extreme constant mass-loss rates of $5 \times 10^{-8}$ 
and $1 \times
10^{-7}\,  M_{\sun}$/yr   along  the   hot  post-AGB  track   from  the
low-gravity domain  at \teff $\approx$ 115000~K sustained  all the way
down to the  hot white dwarf cooling sequence  at about \teff=80000~K.
As a result, we find  that the helium content that eventually survives
in the star amounts to $1.4 \times 10^{-3}$ and $9.4 \times 10^{-4} \,
M_{\sun}$, respectively.  The  total mass lost by the  star amounts to
about $0.013\, M_{\sun}$ in both cases. Most of the mass is lost after
the star  has reached the point  of maximum \teff.  We  note that mass
loss uncovers deep regions in  the star where the chemical composition
varies with depth.   Thus,  the outer layer chemical stratification
after the  end of mass  loss at the  start of the white  dwarf cooling
track looks somewhat  different from the situation in  which mass loss
is not  considered.  In  particular, the external  chemical interfaces
are  markedly smoother.   The implications  for the  chemical profiles
expected during  the DB instability strip are  clearly visualized with
the      help     of     Figs.       \ref{diffusion-lin_2.eps}     and
\ref{diffusion-lin_3.eps} for the mass-loss rates of $5\times 10^{-8}$
and $10^{-7} \, M_{\sun}$/yr,  respectively.  Note in particular that,
in  the  case  of the  lowest  helium  content,  a  single and  not  a
double-layered profile is more appropriate to describe the outer layer
chemical structure of pulsating DB  white dwarfs. This is in line with
the conclusion arrived at in  Althaus \& C\'orsico (2004) that if post
born-again DB white dwarf progenitors are formed with a helium content
smaller than  $10^{-3} \, M_{\sun}$ a double-layered  structure is not
expected  by  the  time the  star  reaches  the  red  edge of  the  DB
instability strip.   Thus, the calculations presented  here place that
conclusion on a  more solid basis.  It is also  clear that the initial
chemical profile is different  according to whether mass loss actually
occurs or not,  an aspect which is expected to  bear some signature on
the pulsational properties of variables DBs.

\begin{figure}
\centering
\includegraphics[clip,width=250pt]{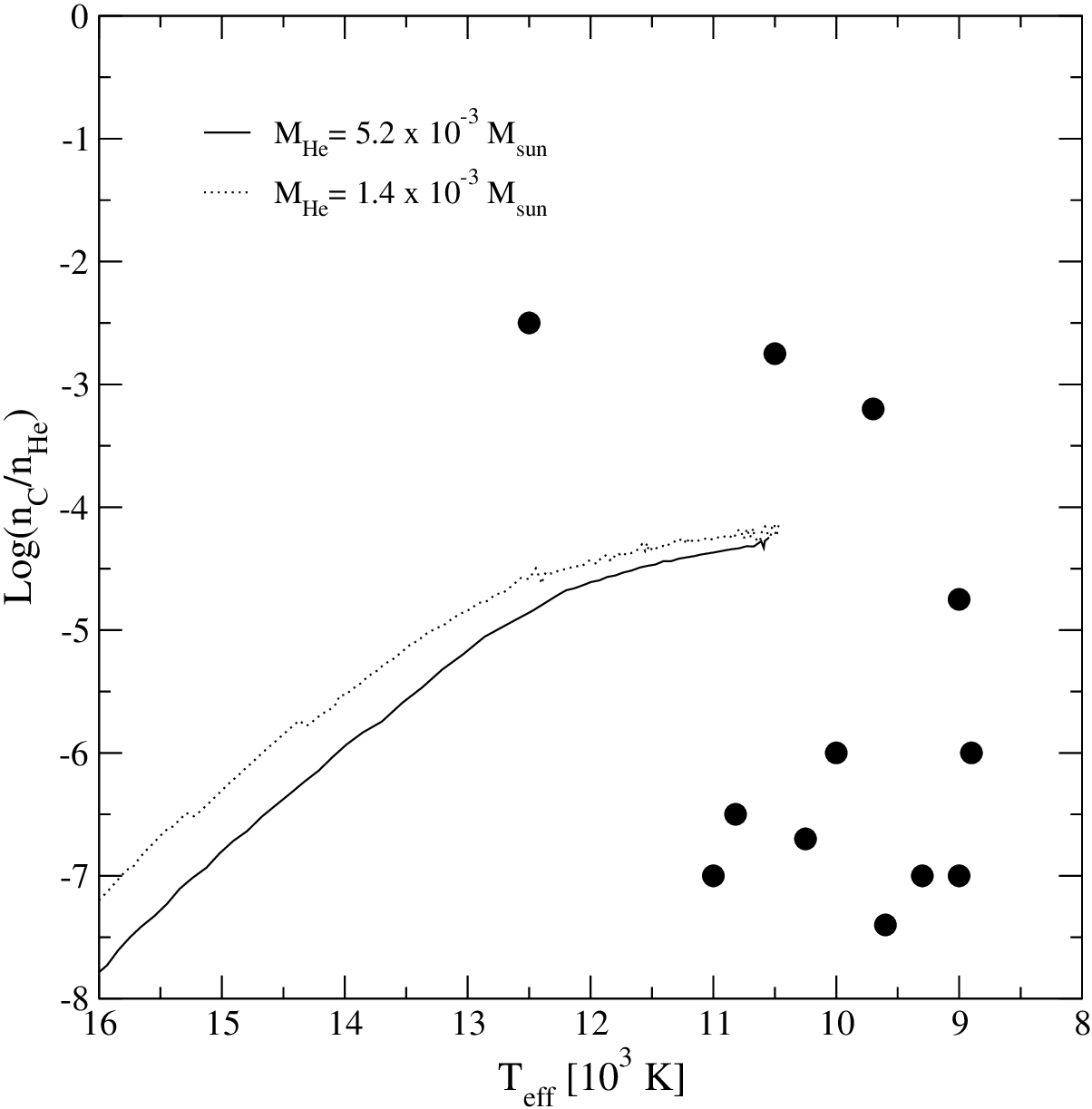}
\caption{Number  density  of  surface $^{12}$C, relative  to  that  of  
$^{4}$He as a function of the effective temperature.  Solid and dotted
lines correspond  to the  white dwarf remnant  with helium  content of
$5.2   \times  10^{-3}$   and  $1.4   \times  10^{-3}   \,  M_{\sun}$,
respectively.  Filled  circles refer  to observed carbon  abundance as
taken from MacDonald et al. (1998).}
\label{ninhe-final.eps}
\end{figure}

We want to mention that, even in the case of extreme mass-loss rates of
$10^{-7}\, M_{\sun}$/yr , the helium envelope is not completely removed. 
However, the rate is
high enough to erode any vestige of $^{14}$N before the remnant reaches 
$T_{\rm eff} \approx 88000$ K. The detection of abundant $^{14}$N 
in some coolest PG1159 stars (for instance in PG0122+200 at 
$T_{\rm eff} \approx 80000$ K; see Dreizler \& Heber 1998) makes the 
persistence of such extreme mass loss rather unlikely.

Additionally,  we investigate the  consequences of  mass-loss episodes
during the hot  post-AGB phase for the surface  abundances expected in
DQ stars.  The predicted  surface $^{12}$C abundance together with the
observed  carbon  abundance  in  DQ  atmospheres  are  shown  in  Fig.
\ref{ninhe-final.eps}. Our  results indicate  that if DQ  white dwarfs
are the  descendants of the  post-born again PG1159 stars,  then their
surface  carbon  abundance  is   not  expected  to  exhibit  a  marked
dependence on the helium content with which the white dwarf is formed.
In  particular, note  that the  carbon abundance  far exceeds  the low
carbon  abundances  detected in  numerous  DQs.   This  prompts us  to
suggest that the DQs with low carbon abundance cannot be linked to the
PG1159 stars if canonical convective dredge-up is the source of carbon
for such DQs.  Instead, they  appear more likely to be good candidates
for an  evolutionary connection  that link them  with stars  that have
somehow  avoided   the  AGB  thermally  pulsing  phase   such  as  the
hydrogen-poor AGB manqu\'e or  the RCrB stars, with normal carbon
abundances. 

\begin{figure}
\centering
\includegraphics[clip,width=250pt]{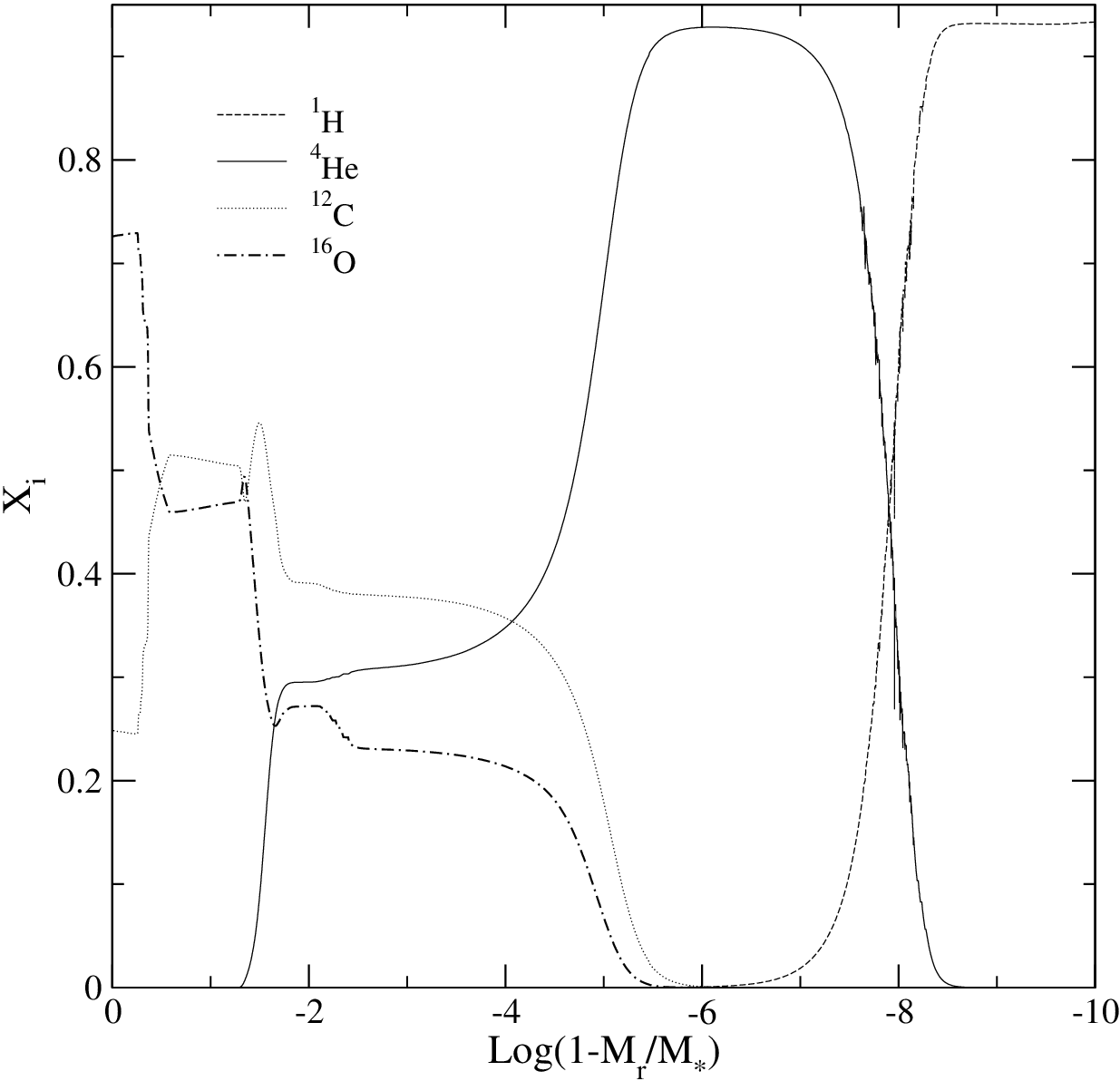}
\caption{Abundance  by   mass  of  $^{1}$H,   $^{4}$He,  $^{12}$C  and
$^{16}$O as a  function of the outer mass fraction  for the $0.5885 \,
M_{\sun}$ white dwarf remnant at \teff= 27000~K. Here, post-born again
mass-loss  episodes  have been  neglected.  Thus,  traces of  hydrogen
remain in the star at the beginning of the white dwarf evolution. Note
that element diffusion leads to the formation of a DA white dwarf with
a  pure  hydrogen  envelope  mass   of  about  $5  \times  10^{-9}  \,
M_{\sun}$.  In  this  particular  simulation, we  have  neglected  the
diffusion of minor species such as $^{13}$C and $^{14}$N.}
\label{diffusion-da.eps}
\end{figure}

Finally, we comment  on the fact that high  mass-loss rates like those
suggested  by  observations  have  consequences for  the  evolutionary
time-scales of the star.  We find that for $\dot{M}= 5 \times 10^{-8}$
and  $10^{-7}  \, M_{\sun}$/yr,  the  star  takes  about $2.25  \times
10^{5}$ and $1.5 \times 10^{5}$~yr, respectively, to evolve from
\teff= 117000~K down to \teff= 81000~K.  This is markedly shorter than 
the time ($4  \times 10^{5}$~yr) needed for the  star to evolve across
the same \teff\ interval in  the absence of mass loss.  This reduction
in the  evolutionary time-scales induced  by mass-loss is  expected to
yield larger  rates of period  change in pulsating GW~Vir  stars, thus
helping to  partially overcome the discrepancy between  theory and the
observed value in the pulsating star PG1159-035 (Costa et al. 1999).
 
In closing  this section, we comment  on the role played  by the small
amount of  hydrogen that survives  the proton burning during  the last
helium thermal  pulse.  As previously mentioned,  the hydrogen content
eventually remaining  in the  star amounts to  $1.3 \times  10^{-8} \,
M_{\sun}$, with traces of this element reaching layers as deep as $3
\times 10^{-4} \, M_{\sun}$  below  the  stellar  surface.  After  the
born-again phase, the total time spent by the remnant in the red-giant
domain amounts  to about  300--400~yr. Consequently, a  mass-loss rate
larger than  $10^{-6} \,  M_{\sun}$/yr would be  enough to  remove, 
during this time interval, the
last vestiges of hydrogen-rich material left in the star. It should be
noted as well  that the development of a  surface lacking any hydrogen
could also  occur during the further  PG1159 evolution as  a result of
hydrogen  burning and  the persistence  of  a constant  wind of  about
$10^{-9}  \,  M_{\sun}$/yr  down   to  the  hot  white  dwarf  cooling
branch. In the results presented thus  far, we have assumed that it is
indeed the actual course of  events.  From the opposite point of view,
however,  that is  in the case of much weaker or less persistent 
mass-loss  events,  we would
expect the  formation of a  DA white dwarf  with a very  thin hydrogen
envelope as a result of element diffusion. This is indeed borne out by
Fig. \ref{diffusion-da.eps}, which  illustrates the chemical abundance
distribution by the  time the white dwarf reaches  \teff= 27000~K.  
The inner hydrogen has been diffused outwards and has formed a pure
hydrogen envelope of mass $5 \times 10^{-9} \, M_{\sun}$, if the extreme
situation of no mass loss is assumed, turning
the white dwarf into one of the DA type. Thus,  
under these circunstances, it is conceivable that  the born-again  
episode could also  give rise  to DA white dwarfs with very thin hydrogen 
envelopes.

\section{Conclusions}

In this paper we have  studied some relevant aspects for the evolution
of hydrogen-deficient white dwarfs  (hereafter referred to as DB white
dwarfs) by  means of new evolutionary  models based on  a complete and
self-consistent  treatment of  the  evolutionary stages  prior to  the
white dwarf  formation.  Specifically, we focused on  DB white dwarfs,
the progenitors  of which have experienced a  born-again episode, that
is a very  late helium thermal pulse on  the early white-dwarf cooling
branch after hydrogen  burning has almost ceased.  The  inclusion of a
time-dependent  scheme  for  the  simultaneous  treatment  of  nuclear
evolution  and mixing processes  due to  convection, salt  fingers and
diffusive overshoot has allowed us  to perform a detailed study of the
abundance   changes  throughout  all   of  the   evolutionary  phases,
particularly during the thermally  pulsing AGB state and the extremely
short-lived phase of the  born-again episode, for which the assumption
of   instantaneous   mixing   becomes  completely   inadequate.    Our
calculations  made use  the double-diffusive  mixing length  theory of
convection  for fluids  with composition  gradients (Grossman  \& Taam
1996).   The study  can thus  be considered  as an  assessment  of its
performance in low-mass stars.  In particular, we have concentrated on
the evolution of an initially $2.7 \, M_{\sun}$ star from the zero-age
main sequence  through the thermally  pulsing and mass-loss  phases to
the white dwarf stage.

As for the  core helium burning and thermally  pulsing AGB phases, our
main results are:

\begin{itemize}

\item The  inner  carbon-oxygen  profile  exhibits  a sharp  variation
      around  $M_{\rm   r}  \approx  0.33  \,   M_{\sun}$  induced  by
mechanical overshoot.

\item Our  models  experience  the  third  dredge-up  during  the AGB. 
      Also, overshooting from below the short-lived  helium-flash convection
      region  at pulse  peak causes  the intershell  region  below the
      helium-rich  buffer  to  be enriched  with  abundant  oxygen, in 
      agreement with Herwig (2000).

\item We find  that pockets of $^{13}$C and $^{14}$N are formed at the 
      base  of  the helium  buffer  after  the  end of  the  dredge-up
      phase. During the  interpulse  period,  the  $^{13}$C-pocket  is
      radiatively  burnt,  whilst the  $^{14}$N-pocket is  engulfed by  
      the helium-flash convection zone during the next thermal pulse.

\end{itemize}

\noindent For the born-again evolution, the following results are worth
mentioning:

\begin{itemize}

\item The  convection  theory of  Grossman \& Taam (1996)  appears  to 
      provide  markedly shorter  born-again time-scales  than previous
      computations;  this  without   invoking  any  reduction  in  the
      convective efficiency.  However, they are still  larger than the
      evolutionary time-scale of  the born-again Sakurai object.  This
      is a  preliminary result, however,  which should be  explored in
      more detail than attempted here.

\item After  the  born-again  phase,  we  find  the  occurrence  of  a 
      double-loop path  in the HR  diagram; that is, the  star reaches
      giant dimensions for the second time after the onset of the last
      helium thermal  pulse and before finally returning  to the white
      dwarf  cooling track. The  amount of  hydrogen remaining  in the
      star is $1.3  \times 10^{-8} \, M_{\sun}$. Most  of the hydrogen
      envelope  burning  occurs in  about  1  month. 

\item After the  born-again episode,  $^{4}$He, $^{12}$C  and $^{16}$O  
      are by far the dominant  surface species with mass abundances of
      ($^{4}$He,$^{12}$C,$^{16}$O)=  (0.306,0.376,0.228).  Amongst the
      main remaining constituents $^{13}$C, $^{14}$N and $^{22}$Ne are
      found   with   mass   fractions   of   4,  1.2   and   2.1   \%,
      respectively. This is in agreement with  surface  abundance patterns
      observed  in  most  hydrogen-deficient  post-AGB stars  such  as
      PG1159  stars. The high  $^{14}$N surface  abundance we  find is
      also in line  with that detected by Dreizler  \& Heber (1998) in
      some PG1159 stars.

\item During  the  early  PG1159  stage,  about  $3 \times  10^{-4} \, 
      M_{\sun}$     of    $^{13}$C     is     processed    via     the
      $^{13}$C$(\alpha,n)^{16}$O reaction. Also, the helium content in
      the star  is reduced  from $8.5 \times  10^{-3}$ to  $5.2 \times
      10^{-3} \, M_{\sun}$ as a result of helium burning.

\end{itemize}

Additionally, the  inclusion of the various  processes responsible for
the  element diffusion  expected  during the  white  dwarf regime  has
enabled us  to extend the scope  of our calculations to  the domain of
the  helium-rich,  carbon  contaminated  DQ  white  dwarfs,  and  more
importantly, to assess the plausibility of the evolutionary connection
PG1159-DB-DQ. In this regard, our main conclusions are:

\begin{itemize}

\item Diffusion  processes lead to  the formation  of a double-layered 
      chemical structure  during the  DB evolution. In  particular, by
      the time the domain of the pulsating  DBs is reached, above  the
      helium-,  carbon- and  oxygen- rich  intershell, there  exists a 
      pure  helium  layer of $ 2.3  \times 10^{-6} \, M_{\sun}$, which 
      increases  to  $1.9  \times  10^{-5} \, M_{\sun}$ when  the star 
      reaches the DQ domain.

\item As  a  result  of convective  dredge-up  at low  \teff,  we find 
      superficial  $^{12}$C  with  abundances  far exceeding  the  low
      carbon abundances  observed in  many DQs.   Our results suggest that
      the DQs with observed low  carbon abundances cannot be linked to
      the PG1159 stars if canonical convective dredge-up is the source
      of  carbon  for such  DQs. 

\end{itemize}

In  closing,  we  believe  that   our  study  would  be  relevant  for
pulsational   applications  since   it   provides  a   self-consistent
description  of the  composition profile  for  both the  core and  the
envelope.  As recently  shown by  Metcalfe et  al. (2003)  this  is an
important  aspect  as  far  as  asteroseismological  inferences  about
pulsating DB white dwarfs  are concerned. Additionally, our new models
constitute a physically sound  frame for answering some open questions
about  pulsating GW~Vir  stars,  the hotter  pulsating predecesors  of
variable DB stars (Gautschy 1997).  We deem that our full evolutionary
models  would  help  to  shed  new light  on  the  physical  processes
responsible for the driving  of pulsations in these stars. Work in
this direction is in progress. 

Detailed tabulations of our post-born again model are freely available
at the following URL: {\tt http://www.fcaglp.unlp.edu.ar/evolgroup/}

\begin{acknowledgements}

We warmly acknowledge  T. Bl\"ocker and K. Werner  for sending us some
reprints central to this work. We also acknowledge A. Gautschy for  a careful  
reading  of the  manuscript. The paper also profited from
a constructive criticism of an anonymous referee.  L.G.A  also
acknowledges the Spanish  MCYT for a Ram\'on y  Cajal Fellowship.
A.M.S. has been supported by the W. M. Keck Foundation through a grant
to the IAS and by the National Science Foundation through the grant
PHY--0070928. Part
of this work  has been supported by the  Instituto de Astrof\'{\i}sica
La Plata, by  the MCYT grant AYA2002--4094--C03--01, by  the CIRIT and
by the European Union FEDER funds.

\end{acknowledgements}


\begin{thebibliography}{}

\bibitem{A94} Alexander, D. R., \& Ferguson, J. W. 1994, ApJ, 437, 879 
\bibitem{AC04} Althaus, L. G., \& C\'orsico, A. H. 2004, A\&A, 417, 1115 
\bibitem{AEA03} Althaus, L. G., Serenelli, A. M., C\'orsico, A. H., \&
        Montgomery, M. H. 2003, A\&A, 404, 593
\bibitem{AG86} Anders, E., \& Grevesse, N. 1986, Geochim. Csomochim. Acta, 
        53, 197
\bibitem{AEA99} Angulo, C., et al. 1999, Nucl. Phys. A, 656, 3 
\bibitem{AT69} Arnett, W. D., \& Truran, J. W. 1969, ApJ, 157, 339 
\bibitem{B75} Bl\"ocker, T., 1995, A\&A, 297, 727 
\bibitem{B01} Bl\"ocker, T. 2001, ApSS, 275, 1 
\bibitem{B69} Burgers,  J. M. 1969,  {\sl ``Flow Equations for Composite 
        Gases''}, Academic, New York
\bibitem{CF88} Caughlan,  G. R., \&  Fowler, W. A. 1988, Atomic  Data
        and Nuclear Data Tables, 40, 290
\bibitem{CEA04} C\'orsico, A. H., Althaus, L. G., Montgomery, M. H., 
        Garc\'{\i}a--Berro, E., \& Isern, J. 2004, A\&A, to be published
\bibitem{CEA99} Costa, J. E. S., Kepler, S. O., Winget, D. E. 1999, ApJ, 
        522, 973 
\bibitem{DK95} Dehner, B. T., \& Kawaler, S. D. 1995, ApJ, 445, L141 
\bibitem{DH98} Dreizler, S., \& Heber, U. 1998, A\&A, 334, 618 
\bibitem{DW96} Dreizler, S., \& Werner, K. 1996, A\&A, 314, 217 
\bibitem{FB02} Fontaine, G., \&  Brassard, P. 2002, ApJ, 581, L33 
\bibitem{F77} Fujimoto, M. Y.  1977, PASJ, 29, 331 
\bibitem{G77} Gautschy, A. 1997, A\&A, 320, 811
\bibitem{GA02} Gautschy, A., \& Althaus, L. G. 2002, A\&A, 382, 141         
\bibitem{GT96} Grossman, S. A., \& Taam, R. E. 1996, MNRAS, 283, 1165 
\bibitem{GNA93} Grossman, S. A., Narayan, R., \& Arnett, D. 1993, ApJ, 
        407, 284 
\bibitem{H00} Herwig, F. 2000, A\&A, 360, 952 
\bibitem{H01} Herwig, F. 2001, ApJ, 554, L71 
\bibitem{H03} Herwig, F. 2003, in {\sl ``Planetary Nebulae: their evolution 
        and role in the  Universe''}, ed. S. Kwok, M. Dopita  and R. 
        Sutherland, IAU symp. 209, ASP Conference Series, pag. 111
\bibitem{HBL99} Herwig, F., Bl\"ocker, T., Langer, N., \& Driebe, T. 1999,
        A\&A, 349, L5
\bibitem{HEA97} Herwig, F., Bl\"ocker, T., Sch\"onberner, D., \&  El Eid, 
        M. 1997, A\&A, 324, L81
\bibitem{IM95} Iben, I. Jr., \& MacDonald, J. 1995, in {\sl ``White Dwarfs''},
        Proc.  of  the  9$^{\rm   th}$  European  Workshop  on  white  dwarfs,
        ed. D. Koester \& K. Werner, Berlin, Springer, 443, 48
\bibitem{IEA83} Iben, I. Jr., Kaler, J. B., Truran, J. W., \&  Renzini, A.
        1983, ApJ, 264, 605
\bibitem{IR96} Iglesias, C. A., \& Rogers, F. J. 1996, ApJ, 464, 943 
\bibitem{I97} Itoh, N., 1997, in {\sl ``Advances in Stellar Evolution''}, 
        Proc. of  the Workshop  ``Stellar Ecology'' (Cambridge,  UK: 
        Cambridge Univ.  Press), Eds.: R.T. Rood \& A. Renzini, p. 185
\bibitem{IEA94} Itoh, N., Hayashi, H., \& Kohyama, Y., 1994, ApJ, 436, 418
\bibitem{KB94} Kawaler, S. D., \& Bradley, P. A. 1994, ApJ, 427, 415 
\bibitem{KW82} Koester, D., Weidemann, V., \& Zeidler-K.T., E. M. 1982, A\&A, 
        116, 147
\bibitem{KH97} Koesterke, L., \& Hamann, W. R. 1997, A\&A, 320, 91 
\bibitem{KW98} Koesterke, L., \& Werner, K. 1998, ApJ, 500, L55 
\bibitem{LM03} Lawlor, T. M., \& MacDonald, J. 2003, ApJ, 583, 913 
\bibitem{LEA03} Lugaro, M., Herwig, F., Lattanzio, J. C., Gallino, R., 
        Straniero, O. 2003, ApJ, 586, 1305
\bibitem{MEA98} MacDonald, J., Hernanz, M., \& Jos\'e, J. 1998, MNRAS, 
        296, 523 
\bibitem{MM79} Magni, G., \& Mazzitelli, I. 1979, A\&A, 72, 134 
\bibitem{MEA99} Mazzitelli, I., D'Antona, F., \& Ventura, P. 1999, A\&A, 
        348, 846 
\bibitem{MEA03} Metcalfe, T. S., Montgomery, M. H., \&  Kawaler, S. D. 2003, 
        MNRAS, 344, L88
\bibitem{OK00} O'Brien, M. S., \& Kawaler, S. D. 2000, ApJ, 539, 372  
\bibitem{OEA98} O'Brien, M. S., et al. 1998, ApJ, 495, 458 
\bibitem{PEA86} Pelletier,  C., Fontaine,  G., Wesemael, F., Michaud, G., 
        \& Wegner, G. 1986, ApJ, 307, 242
\bibitem{S79} Sch\"onberner, D. 1979, A\&A, 79, 108 
\bibitem{S96} Sch\"onberner, D. 1996, in {\sl ``Hydrogen-Deficient Stars''},
        ed. C. S. Jeffery and U. Heber, ASP Conference Series, vol. 96, 433
\bibitem{SH65} Schwarzschild, M.,  \& H\"arm, R.  1965, ApJ, 142, 855 
\bibitem{SEA03} Straniero, O., Dom\'{\i}nguez, I., Imbriani, G., \& Piersanti, 
        L. 2003, ApJ, 583, 878
\bibitem{UB00} Unglaub, K., \& Bues, I. 2000, A\&A, 359, 1042 
\bibitem{VEA99} Ventura, P., D'Antona, F., \& Mazzitelli, I. 1999, ApJ, 
        524, L111 
\bibitem{WK95} Weidemann, V., \&  Koester, D. 1995, A\&A, 297, 216 
\bibitem{W01} Werner, K. 2001, Ap\&SS, 275, 27 
\bibitem{WEA97} Werner, K., Dreizler, S., Heber, U., Kappelmann, N., Kruk, J., 
        Rauch, T., \& Wolff, B. 1997, Reviews of Modern Astronomy, 10, 219
\bibitem{WEA04} Winget, D. E., Sullivan, D. J., Metcalfe, T. S., Kawaler, 
        S. D., \& Montgomery, M. H. 2004, ApJ, 602, L109

\end{thebibliography}
\end{document}